\documentclass[11pt,a4paper]{article}
\pdfoutput=1 

\usepackage[table]{xcolor}
\usepackage{colortbl}
\usepackage{comment}
\RequirePackage{ifpdf} 
\usepackage{amsmath} 
\usepackage{mathtools}

\usepackage{jheppub}
\usepackage{pstricks}
\usepackage[final]{pdfpages} 
\usepackage{ifpdf} 
\usepackage{slashed}
\usepackage{pdflscape}
\usepackage[normalem]{ulem}
\usepackage{color}
\usepackage{xcolor}
\definecolor{urlblue}{rgb}{0.2,0.4,0.7} 
\definecolor{citegreen}{rgb}{0,0.6,0.2}
\definecolor{linkred}{rgb}{0.9,0.2,0.1}
\usepackage{hyperref}
\hypersetup{
colorlinks=true, citecolor=citegreen, linkcolor=blue, urlcolor=urlblue}

\usepackage{graphics}
\usepackage{etoolbox} 
\usepackage{fixmath}
\usepackage{psfrag}

\usepackage{notoccite} 

\usepackage{amsfonts}
\usepackage{autobreak}
\usepackage{marginnote}
\usepackage{enumitem}
\usepackage{appendix}
\usepackage{caption}
\usepackage{tabularx}
\usepackage{graphicx}
\newcommand{\NOdisplay}[1]{ }

\newcommand{\eg}{{\it e.g.}}
\newcommand{\ie}{{\it i.e.}}
\newcommand{\derive}{{\mathrm d}}
\newcommand{\dlog}{\mathrm{dlog}}

\def\qgraf{{\sc
Qgraf}}
\def\feynarts{{\sc
FeynArts}}

\def\kira{{\sc
Kira}}
\def\finiteflow{{\sc
FiniteFlow}}
\def\litered{{\sc
LiteRed}}

\def\superchic{{\sc
SuperChic}}
\def\gammaupc{{\sc
gamma-UPC}}
\def\helaconia{{\sc
HELAC-Onia}}
\def\mg{{\sc
MadGraph5\_aMC@NLO}}
\def\fastgpl{{\sc FastGPL}}
\def\handyg{{\sc HandyG}}
\def\amflow{{\sc AMFlow}}
\def\madloop{{\sc MadLoop}}
\def\multivariateapart{{\sc MultivariateApart}}

\definecolor{mypink}{RGB}{219, 48, 122}

\usepackage{tikz}
\usetikzlibrary{positioning,arrows}
\usetikzlibrary{decorations.pathmorphing}
\usetikzlibrary{decorations.markings}
\usetikzlibrary{shapes.geometric}
\tikzset{
    vector/.style={decorate, decoration={snake}, draw},
    provector/.style={decorate, decoration={snake,amplitude=2.5pt}, draw},
    antivector/.style={decorate, decoration={snake,amplitude=-2.5pt}, draw},
    fermion/.style={draw=black,
      postaction={decorate},decoration={markings,mark=at position .55
        with {\arrow[draw=black]{>}}}},
    fermionbar/.style={draw=black, postaction={decorate},
                       decoration={markings,mark=at position .55 with {\arrow[draw=black]{<}}}},
    fermionnoarrow/.style={draw=black},
    gluon/.style={decorate, draw=red,decoration={coil,amplitude=4pt, segment length=6pt}},
    photon/.style={decorate, draw=red,decoration={snake,amplitude=3pt, segment length=6pt}},
    scalar/.style={dashed,draw=black,
      postaction={decorate},decoration={markings,mark=at position .55
        with {\arrow[draw=black]{>}}}},
    scalarbar/.style={dashed,draw=black,
      postaction={decorate},decoration={markings,mark=at position .55
        with {\arrow[draw=black]{<}}}},
    scalarnoarrow/.style={dashed,draw=black},
    electron/.style={draw=black,
      postaction={decorate},decoration={markings,mark=at position .55
        with {\arrow[draw=black]{>}}}},
    bigvector/.style={decorate, decoration={snake,amplitude=4pt}, draw},
}

\title{Two-loop massive QCD and QED helicity amplitudes for light-by-light scattering}

\author{Ajjath~A~H$^{a}$, Ekta Chaubey$^{b}$, Hua-Sheng Shao$^{a}$}
\emailAdd{aabdulhameed@lpthe.jussieu.fr, eekta@uni-bonn.de, huasheng.shao@lpthe.jussieu.fr}
\affiliation{$^a$Laboratoire de Physique Théorique et Hautes Energies (LPTHE), UMR 7589, Sorbonne Université et
CNRS, 4 place Jussieu, 75252 Paris Cedex 05, France
\\$^b$Bethe Center for Theoretical Physics, Universität Bonn, 53115 Bonn, Germany
}

\preprint{}

\abstract{We present the analytic and compact two-loop helicity amplitudes for QCD and QED corrections to the light-by-light scattering process with massive internal fermions. We express the master integrals either in terms of multiple polylogarithms or in terms of iterated integrals with $\dlog$ one-forms.  We also elaborate on optimising the analytic results for each phase-space region. This makes the numerical evaluation of the scattering amplitudes fast, stable and suitable for phenomenological applications. }

\begin{document}
\allowdisplaybreaks[4]
\unitlength1cm
\keywords{}
\maketitle
\flushbottom

\section{Introduction}
\label{sec:intro}
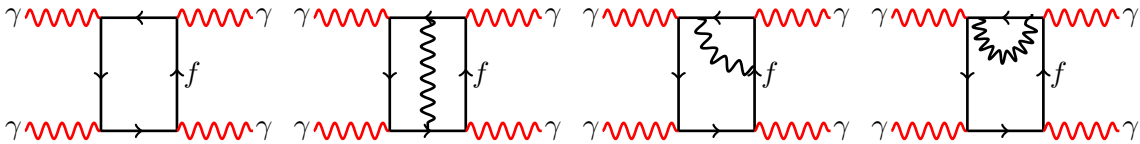
\begin{figure}[h!]
  \begin{tikzpicture}[line width=1 pt, scale=0.5]
  \hspace{0cm}
\draw[photon] (-2,1.5) -- (0,1.5);
\draw[photon] (-2,-1.5) -- (0,-1.5);
\draw[fermion] (0,1.5) -- (0,-1.5);
\draw[fermion] (0,-1.5) -- (2,-1.5);
\draw[fermion] (2,-1.5) -- (2,1.5);
\draw[fermion] (2,1.5) -- (0,1.5);
\draw[photon] (2,1.5) -- (4,1.5);
\draw[photon] (2,-1.5) -- (4,-1.5);
\node at (-2.3,1.5) {$\gamma$};
\node at (-2.3,-1.5) {$\gamma$};
\node at (4.3,1.5) {$\gamma$};
\node at (4.3,-1.5) {$\gamma$};
\node at (2.4,0) {$f$};
\hspace{3.8cm}
\draw[photon] (-2,1.5) -- (0,1.5);
\draw[photon] (-2,-1.5) -- (0,-1.5);
\draw[vector, segment length=6pt] (1,1.5) -- (1,-1.5);
\draw[fermion] (0,1.5) -- (0,-1.5);
\draw[fermion] (0,-1.5) -- (2,-1.5);
\draw[fermion] (2,-1.5) -- (2,1.5);
\draw[fermion] (2,1.5) -- (0,1.5);
\draw[photon] (2,1.5) -- (4,1.5);
\draw[photon] (2,-1.5) -- (4,-1.5);
\node at (-2.3,1.5) {$\gamma$};
\node at (-2.3,-1.5) {$\gamma$};
\node at (4.3,1.5) {$\gamma$};
\node at (4.3,-1.5) {$\gamma$};
\node at (2.4,0) {$f$};
\hspace{3.8cm}
\draw[photon] (-2,1.5) -- (0,1.5);
\draw[photon] (-2,-1.5) -- (0,-1.5);
\draw[vector, segment length=6pt, bend right,looseness = 1.5] (0.5,1.5) to (2,0.2);
\draw[fermion] (0,1.5) -- (0,-1.5);
\draw[fermion] (0,-1.5) -- (2,-1.5);
\draw[fermion] (2,-1.5) -- (2,1.5);
\draw[fermion] (2,1.5) -- (0,1.5);
\draw[photon] (2,1.5) -- (4,1.5);
\draw[photon] (2,-1.5) -- (4,-1.5);
\node at (-2.3,1.5) {$\gamma$};
\node at (-2.3,-1.5) {$\gamma$};
\node at (4.3,1.5) {$\gamma$};
\node at (4.3,-1.5) {$\gamma$};
\node at (2.4,0) {$f$};
\hspace{3.8cm}
\draw[photon] (-2,1.5) -- (0,1.5);
\draw[photon] (-2,-1.5) -- (0,-1.5);
\draw[black,decorate,decoration={snake,amplitude=2.3pt, segment length=4pt
}, out=-90, in =-90,looseness=2.5] (0.3,1.5) to (1.7,1.6);
\draw[fermion] (0,1.5) -- (0,-1.5);
\draw[fermion] (0,-1.5) -- (2,-1.5);
\draw[fermion] (2,-1.5) -- (2,1.5);
\draw[fermion] (2,1.5) -- (0,1.5);
\draw[photon] (2,1.5) -- (4,1.5);
\draw[photon] (2,-1.5) -- (4,-1.5);
\node at (-2.3,1.5) {$\gamma$};
\node at (-2.3,-1.5) {$\gamma$};
\node at (4.3,1.5) {$\gamma$};
\node at (4.3,-1.5) {$\gamma$};
\node at (2.4,0) {$f$};
\end{tikzpicture}
\caption{Representative Feynman diagrams of $\gamma\gamma\to \gamma\gamma$ at one loop (left) and two loops (rest). The internal wavy line can be either a gluon or a photon.} 
\label{fig:feynmandiag}
\end{figure}

Light-by-light (LbL) scattering is a fundamental quantum electrodynamics (QED) process resulting from the quantum effects predicted in the 1930s~\cite{Heisenberg:1934pza,Euler:1935zz,Euler:1935qgl,Heisenberg:1936nmg}. It is a loop-induced process via one-loop Feynman diagrams with charged internal particles in the loop, as illustrated in the left panel of figure~\ref{fig:feynmandiag}. The lowest order of cross section is $\mathcal{O}(\alpha^4)$, where $\alpha$ is the electromagnetic fine structure constant. In the low energy limit, the process can be effectively described by the Euler-Heisenberg Lagrangian~\cite{Heisenberg:1936nmg}. The first complete calculation in QED was carried out by Karplus and Neuman~\cite{Karplus:1950zz} in 1951. The interest in studying LbL revived recently, in particular from the point-of-view of searching for the elusive beyond the Standard Model (BSM) signals. It has been acknowledged that LbL can be used to probe the
quartic anomalous gauge couplings~\cite{dEnterria:2013zqi}, the axion-like particles~\cite{Knapen:2016moh}, the graviton-like particles~\cite{dEnterria:2023npy,Atag:2010bh}, the nonlinear Born-Infeld extension of QED as well as in the context of the Standard Model (SM)~\cite{Ellis:2017edi}, the photon-photon self-interaction from the noncommutative QED~\cite{Horvat:2020ycy}, large extra dimensions~\cite{Cheung:1999ja,Davoudiasl:1999di}, and supersymmetric particles~\cite{Greiner:1992fz}. LbL is also a background for looking for new particles in the SM, such as ditauonium~\cite{dEnterria:2022ysg,dEnterria:2023yao} and glueballs~\cite{Greiner:1992fz}.

In spite of the early theoretical proposals, the \textit{direct} experimental observations~\footnote{LbL has also been \textit{indirectly} tested by the precise measurements of the anomalous magnetic moments of the electron and the muon and by the elastic scattering of a photon in Coulomb fields of nuclei via the so-called Delbr\"uck scattering.} of the LbL scattering process $\gamma\gamma\to\gamma\gamma$ were only achieved recently in the ultraperipheral heavy-ion collisions (UPCs) at the LHC~\cite{ATLAS:2017fur,CMS:2018erd,ATLAS:2019azn,ATLAS:2020hii}. Its experimental feasibility, first suggested in ref.~\cite{dEnterria:2013zqi}, can be mostly attributed to the large coherent photon flux carried by the ultra-relativistic nucleus thanks to its large charge number $Z$. For instance, the lead (Pb) beam used by the LHC has $Z=82$. With UPCs, we have the largest magnitude of electromagnetic fields produced in the laboratory in an experimentally controlled way. It enhances the LbL cross section by $Z^4\approx 4.5\cdot 10^7$ in Pb-Pb collisions with respect to the counterparts in proton-proton ($pp$) and electron-positron ($e^-e^+$) collisions. In a UPC exclusive process, the two initial beam hadrons do not break, which can be identified with detectors at very forward angles, Roman Pots and Zero Degree Calorimeters for proton-proton and nucleus-nucleus collisions respectively. This usually provides us with distinct signatures compared to the central and peripheral collisions. The UPC exclusive processes have the virtue of low backgrounds due to their much smaller particle multiplicities than in the inclusive scatterings, and thus the experimental triggers can be loosened. This allows us to probe a different kinematic regime with respect to the inclusive reactions that breakup the beam hadrons.

The LHC measurements have been compared with the theoretical calculations by the two Monte Carlo event generators: \superchic~\cite{Harland-Lang:2020veo} and \gammaupc~\cite{Shao:2022cly}~\footnote{Strictly speaking, \gammaupc\ is a library for calculating the coherent photon-photon flux in UPC. It has been integrated into the two event generators \helaconia~\cite{Shao:2012iz,Shao:2015vga} and \mg~\cite{Alwall:2014hca} to enable the event generations of exclusive UPC processes.}. Both of them are based on the leading order (\ie, one-loop) matrix element and yield the consistent LbL cross sections. However, the theoretical cross sections are about 2 standard deviations below the ATLAS measured value. It is therefore interesting to investigate the origin of this discrepancy. On the theory side, it would be a natural step~\footnote{Another possibly important missing effect is the higher order Coulomb $\mathcal{O}(Z\alpha)$ corrections, which is however beyond the scope of our paper.} to check the size of the next-to-leading order (NLO) quantum corrections that involve two-loop helicity amplitudes. In the low-energy limit, the two-loop corrections to the Euler-Heisenberg Lagrangian have been computed with the string-inspired approach~\cite{Reuter:1996zm,Fliegner:1997ra,Schubert:2001he,Martin:2003gb}. On the other hand, in the high energy limit, two-loop amplitudes with massless internal fermions in QCD and QED~\cite{Bern:2001dg} and in supersymmetric QED~\cite{Binoth:2002xg} have also been available for two decades. The aim of this paper is to present the analytic two-loop helicity amplitudes with general massive internal fermions in the theories of QCD and QED, where some representative graphs are shown in figure~\ref{fig:feynmandiag}. This will allow us to lift the above two approximations, which will also be relevant for the theory-experiment comparison. We invest efforts in simplifying the initial lengthy expressions into more compact and concise forms. This is not only for the purpose of aesthetics, but also to improve the numerical stability of helicity amplitudes at two loops. This optimisation becomes particularly important when dealing with the substantial cancellations between individual Feynman graphs as a consequence of the gauge invariance and infrared (IR) finiteness. The corresponding two-loop Feynman integrals with massive internal lines have also been studied previously in refs.~\cite{Caron-Huot:2014lda,Xu:2018eos,Mandal:2018cdj,Maltoni:2018zvp,Wang:2020nnr}, though the results are only partially available in the literature.



The rest of the paper is organised as follows. We derive the general structure of scattering amplitude to any loop order in section \ref{sec:amp}. The two-loop master integrals are discussed in section \ref{sec:masterintegrals2L}. The analytic expressions of the one- and two-loop helicity amplitudes with full fermion mass dependence are presented in section \ref{sec:analytichelamp}. We draw our conclusion in section \ref{sec:conclusion}. The appendices collect the additional information. We define the kinematic regions of our rationalised variables  as well as the analytic continuation in appendix \ref{sec:wzdef}. The analytic boundary conditions for the non-rationalised two-loop master integrals are given in appendix \ref{app:BCs}. We define the one-loop master integrals in appendix \ref{sec:masterintegrals1L}, and present the explicit expressions of our chosen canonical basis of two-loop master integrals in appendix \ref{sec:canonmasterintegrals2L}. Finally, we report the one-loop helicity amplitudes of the $W^\pm$ boson contribution in appendix \ref{sec:oneloopamp4W}, and collect the one- and two-loop helicity amplitudes in the low-energy limit in appendix \ref{sec:ampLElimit}.

\section{General structure of scattering amplitude}
\label{sec:amp}
In this section, we derive the independent form factors and helicity amplitudes. A similar derivation exists in the literature~\cite{Binoth:2002xg}. In addition, an alternative approach of obtaining helicity amplitudes is also described in ref.~\cite{Peraro:2020sfm}. However, our derivation is neither (completely) identical to ref.~\cite{Binoth:2002xg} nor following ref.~\cite{Peraro:2020sfm}. Therefore, for completeness, we present our derivation in detail in this section.

Let us consider the process
\begin{equation}
\gamma (p_1, \lambda_1) +\gamma (p_2 , \lambda_2) +\gamma (p_3 , \lambda_3) + \gamma (p_4, \lambda_4) \rightarrow 0,
\end{equation}
where all the momenta $p_i$ of the photons are incoming and their helicities are denoted as $\lambda_i$. The amplitude is expressed as
\begin{equation}
\mathcal{M}_{\lambda_1 \lambda_2\lambda_3\lambda_4}= \left(\prod_{i=1}^{4}{\varepsilon_{\lambda_i,\mu_i}(p_i)} \right)\mathcal{M}^{\mu_1 \mu_2 \mu_3 \mu_4}(p_1, p_2, p_3, p_4)
\end{equation}
with $\varepsilon_{\lambda_i,\mu_i}(p_i)$ being the polarisation vectors of external photons. The Lorentz decomposition of the scattering tensor $\mathcal{M}^{\mu_1 \mu_2 \mu_3 \mu_4}$ is given as:~\footnote{The tensorial structure needs to be extended with the Levi-Civita tensor when considering the chiral-dependent electroweak corrections.}
\begin{align}
\mathcal{M}^{\mu_1 \mu_2 \mu_3 \mu_4} =& A_1 g^{\mu_1 \mu_2} g^{\mu_3 \mu_4} +A_2 g^{\mu_1 \mu_3} g^{\mu_2 \mu_4} + A_3 g^{\mu_1 \mu_4} g^{\mu_2 \mu_3} \nonumber\\&+ \sum_{j_1, j_2=1}^3 (B_{j_1 j_2}^1 g^{\mu_1 \mu_2} p_{j_1}^{\mu_3} p_{j_2}^{\mu_4} + B^2_{j_1j_2} g^{\mu_1 \mu_3} p_{j_1}^{\mu_2} p_{j_2}^{\mu_4} +B^3_{j_1j_2} g^{\mu_1 \mu_4} p_{j_1}^{\mu_2} p_{j_2}^{\mu_3} \nonumber\\&+B^4_{j_1j_2} g^{\mu_2 \mu_3} p_{j_1}^{\mu_1} p_{j_2}^{\mu_4} +B^5_{j_1j_2} g^{\mu_2 \mu_4} p_{j_1}^{\mu_1} p_{j_2}^{\mu_3}+B^6_{j_1j_2} g^{\mu_3 \mu_4} p_{j_1}^{\mu_1} p_{j_2}^{\mu_2}) \nonumber\\&+\sum_{j_1, j_2, j_3, j_4=1}^3 C_{j_1 j_2 j_3 j_4} p_{j_1}^{\mu_1} p_{j_2}^{\mu_2} p_{j_3}^{\mu_3} p_{j_4}^{\mu_4}. \label{eq:FFgeneral}
\end{align}
The coefficients $A_i$, $B_{jk}^i$ and $C_{ijkl}$ are functions of the Mandelstam variables $s=(p_1+p_2)^2,\; t=(p_2+p_3)^2,\; u=(p_1+p_3)^2$, as well as the masses of the particles in loops. Therefore, they should be understood as $A_i(s,t,u),B_{jk}^i(s,t,u),C_{ijkl}(s,t,u)$.
The on-shell conditions and the conservation of momenta guarantee that the sum $s+t+u=0$. Thanks to the parity invariance in QCD and QED, the term proportional to the Levi-Civita tensor is forbidden. We have $3+6\times 3\times 3+3^4=138$ independent functions in \eqref{eq:FFgeneral}. However, many of the coefficients are related by the following three symmetries:
\begin{enumerate}
\item \textbf{Transversality:} $\varepsilon_{\lambda_i}\cdot p_i=0$ for $i=1,2,3,4$. This reduces to 57 functions to be considered.
\item \textbf{Bose symmetry:} the tensor amplitude $\mathcal{M}^{\mu_1 \mu_2 \mu_3 \mu_4}$ is invariant under the exchange of any two index pair $(p_i,\mu_i)$ with $i=1,2,3,4$, which gives us $7$ independent functions.
\item \textbf{Gauge symmetry:} the gauge symmetry guarantees the following four Ward identities
\begin{eqnarray}
0&=&\mathcal{M}^{p_1 \varepsilon_{\lambda_2}(p_2) \varepsilon_{\lambda_3}(p_3)\varepsilon_{\lambda_4}(p_4)}=\mathcal{M}^{\varepsilon_{\lambda_1}(p_1)p_2\varepsilon_{\lambda_3}(p_3)\varepsilon_{\lambda_4}(p_4)}\nonumber\\
&=&\mathcal{M}^{\varepsilon_{\lambda_1}(p_1)\varepsilon_{\lambda_2}(p_2)p_3\varepsilon_{\lambda_4}(p_4)}=\mathcal{M}^{\varepsilon_{\lambda_1}(p_1)\varepsilon_{\lambda_2}(p_2)\varepsilon_{\lambda_3}(p_3)p_4}.
\end{eqnarray}
They lead to a few more identities to reduce to the 3 independent functions, which are
\begin{equation}
A_1, \quad \Delta B_{11}^1=B_{11}^1-B_{12}^1, \quad \Delta C_{2111}=C_{2111}-C_{2112}.
\end{equation}
\end{enumerate}
 Our derivation and all the relations have been cross checked with explicit calculations. The number of independent form factors (not functions) in $\mathcal{M}^{\mu_1 \mu_2 \mu_3 \mu_4}$ is $7$, instead of $15$ in section 3.1 of ref.~\cite{Binoth:2002xg}. They are $A_1(s,t,u),\; A_1(t,s,u),\; A_1(u,s,t)$, $\Delta B_{11}^1(s,t,u),\; \Delta B_{11}^1(t,s,u),$ $\Delta B_{11}^1(u,s,t),$ and $\Delta C_{2111}(s,t,u)$.

The reduced tensor structure becomes
\begin{eqnarray}
\mathcal{M}^{\mu_1 \mu_2 \mu_3 \mu_4}&=&A^{\mu_1\mu_2\mu_3\mu_4}+B^{\mu_1\mu_2\mu_3\mu_4}+C^{\mu_1\mu_2\mu_3\mu_4},
\end{eqnarray}
where
\begin{eqnarray}
A^{\mu_1\mu_2\mu_3\mu_4}&=&A_1(s,t,u)\left(g^{\mu_1\mu_2}-\frac{p_1^{\mu_2}p_2^{\mu_1}}{p_1\cdot p_2}\right)\left(g^{\mu_3\mu_4}+\frac{p_1^{\mu_3}+p_2^{\mu_3}}{p_1\cdot p_2}p_3^{\mu_4}\right)\label{eq:Atensor}\\
&&+[(\mu_1,p_1)\leftrightarrow (\mu_3,p_3), s\leftrightarrow t]\nonumber\\
&&+[(\mu_1,p_1)\to (\mu_3,p_3), (\mu_2,p_2)\to (\mu_1,p_1), (\mu_3,p_3)\to (\mu_2,p_2), s\to u, t\to s, u\to t],\nonumber\\
B^{\mu_1\mu_2\mu_3\mu_4}&=&\Delta B_{11}^1(s,t,u)\left[-\left(g^{\mu_1\mu_2}-\frac{p_1^{\mu_2}p_2^{\mu_1}}{p_1\cdot p_2}\right)\left(p_1^{\mu_3}-\frac{u}{t}p_2^{\mu_3}\right)\left(p_2^{\mu_4}-\frac{u}{s}p_3^{\mu_4}\right)\right.\nonumber\\
&&\left.+\left(p_3^{\mu_1}-\frac{u}{s}p_2^{\mu_1}\right)\left(p_1^{\mu_2}-\frac{s}{t}p_3^{\mu_2}\right)\left(g^{\mu_3\mu_4}+\frac{2p_2^{\mu_3}p_1^{\mu_4}}{t}+\frac{2p_1^{\mu_3}p_2^{\mu_4}}{u}\right)\right]\label{eq:Btensor}\\
&&+[(\mu_1,p_1)\leftrightarrow (\mu_3,p_3), s\leftrightarrow t]\nonumber\\
&&+[(\mu_1,p_1)\to (\mu_3,p_3), (\mu_2,p_2)\to (\mu_1,p_1), (\mu_3,p_3)\to (\mu_2,p_2), s\to u, t\to s, u\to t],\nonumber\\
C^{\mu_1\mu_2\mu_3\mu_4}&=&\Delta C_{2111}(s,t,u)\frac{1}{st^2u}\left(sp_3^{\mu_1}-up_2^{\mu_1}\right)\left(sp_3^{\mu_2}-t p_1^{\mu_2}\right)\left(up_2^{\mu_3}-tp_1^{\mu_3}\right)\left(sp_2^{\mu_4}-up_3^{\mu_4}\right).\label{eq:Ctensor}
\end{eqnarray}
In the above eqs.\eqref{eq:Atensor},\eqref{eq:Btensor}, we have used the short-hands for the replacement rules ``$\to$" and the exchange rules ``$\leftrightarrow$" with respect to the first terms on the right-hand sides.
The tensor $C^{\mu_1\mu_2\mu_3\mu_4}$ is invariant under the exchange $[(\mu_1,p_1)\leftrightarrow (\mu_3,p_3), s\leftrightarrow t]$ and the replacement $[(\mu_1,p_1)\to (\mu_3,p_3), (\mu_2,p_2)\to (\mu_1,p_1), (\mu_3,p_3)\to (\mu_2,p_2), s\to u, t\to s, u\to t]$.


In the next step, we need to define the projection operators to obtain the coefficients $A_1$, $\Delta B_{11}^1$ and $\Delta C_{2111}$. We can use the following tensor
\begin{align}
\mathcal{P}^{\mu \nu} &= g^{\mu \nu} - \sum_{j=1}^3{\sum_{k=1}^3{p_j ^\mu \left[\mathbb{G}^{-1}\right]_{jk} p_k^\nu}}, \quad \mathcal{R}_j^{\nu} = \sum_{k=1}^3{\left[\mathbb{G}^{-1}\right]_{jk} p_k^\nu},
\end{align}
where $\mathbb{G}=\mathbb{G}(p_1,p_2,p_3)=\left(\begin{array}{ccc}p_1^2 & p_1\cdot p_2 & p_1\cdot p_3 \\
p_1\cdot p_2 & p_2^2 & p_2\cdot p_3\\
p_1\cdot p_3 & p_2\cdot p_3 & p_3^2\\
\end{array}\right)$ is the Gram matrix.
$\mathcal{P}$ is the projector that projects onto the $(d-3)$-dimensional subspace perpendicular to the space spanned by the external momenta (\ie, $p_{i,\mu}\mathcal{P}^{\mu\nu}=\mathcal{P}^{\mu\nu}p_{i,\nu}=0$). $\mathcal{R}_{j,\nu}$ is the dual vector to $p_j^\nu$ relative to this space (\ie, $\mathcal{R}_{j,\nu}p_i^\nu=\delta_{ji}$). In addition, we have the following relations
\begin{eqnarray}
\mathcal{P}^{\mu}_{\mu}&=&d-3,\nonumber\\
\mathcal{P}^{\mu\rho}\mathcal{P}_{\rho}^{\nu}&=&g^{\mu\nu}-2\sum_{j=1}^3{\sum_{k=1}^3{p_j ^\mu \left[\mathbb{G}^{-1}\right]_{jk} p_k^\nu}}+\sum_{j,k,l,n=1}^{3}{p_j^{\mu}\left[\mathbb{G}^{-1}\right]_{jk} p_{k}^{\rho}p_{l,\rho}\left[\mathbb{G}^{-1}\right]_{ln}p_{n}^{\nu}}\nonumber\\
&=&g^{\mu\nu}-2\sum_{j=1}^3{\sum_{k=1}^3{p_j ^\mu \left[\mathbb{G}^{-1}\right]_{jk} p_k^\nu}}+\sum_{j,k,l,n=1}^{3}{p_j^{\mu}\left[\mathbb{G}^{-1}\right]_{jk} \left[\mathbb{G}\right]_{kl}\left[\mathbb{G}^{-1}\right]_{ln}p_{n}^{\nu}}\nonumber\\
&=&\mathcal{P}^{\mu\nu},\nonumber\\
\mathcal{R}_{i}^{\mu}\mathcal{R}_{j,\mu}&=&\sum_{k,l=1}^{3}{\left[\mathbb{G}^{-1}\right]_{ik} p_k^\mu \left[\mathbb{G}^{-1}\right]_{jl} p_{l,\mu}}=\left[\mathbb{G}^{-1}\right]_{ji}=\left[\mathbb{G}^{-1}\right]_{ij},\nonumber\\
\mathcal{P}^{\mu\nu}\mathcal{R}_{j,\nu}&=&\mathcal{R}_{j}^{\mu}-\sum_{k,l,n=1}^{3}{p_k^{\mu}\left[\mathbb{G}^{-1}\right]_{kl}p_l^{\nu}\left[\mathbb{G}^{-1}\right]_{jn}p_{n,\nu}}=0=\mathcal{R}_{j,\mu}\mathcal{P}^{\mu\nu},
\end{eqnarray}
where we have denoted the space-time dimension by $d=4-2\epsilon$ with $\epsilon$ being the dimensional regularisation parameter that ought to be taken infinitesimal at the end.

We can obtain the $A_i$ tensor coefficients using the following operations
\begin{align}
\tilde{A}_1 &= \frac{1}{(d-1)(d-3)} \mathcal{P}_{\mu_1 \mu_2} \mathcal{P}_{\mu_3 \mu_4} \mathcal{M}^{\mu_1 \mu_2 \mu_3 \mu_4}=\frac{1}{(d-1)}\left[(d-3)A_1+A_2+A_3\right],\\
\tilde{A}_2&= \frac{1}{(d-1)(d-3)}\mathcal{P}_{\mu_1 \mu_3} \mathcal{P}_{\mu_2 \mu_4} \mathcal{M}^{\mu_1 \mu_2 \mu_3 \mu_4}=\frac{1}{(d-1)}\left[A_1+(d-3)A_2+A_3\right],\\
\tilde{A}_3&= \frac{1}{(d-1)(d-3)}\mathcal{P}_{\mu_1 \mu_4} \mathcal{P}_{\mu_2 \mu_3} \mathcal{M}^{\mu_1 \mu_2 \mu_3 \mu_4}=\frac{1}{(d-1)}\left[A_1+A_2+(d-3)A_3\right].
\end{align}
In other words, we have
\begin{align}
\begin{pmatrix}
\tilde{A}_1\\
\tilde{A}_2\\
\tilde{A}_3
\end{pmatrix}=\frac{1}{(d-1)}  \begin{pmatrix}
d-3 & 1 & 1\\
1 & d-3 & 1\\
1 & 1 & d-3
\end{pmatrix} \begin{pmatrix}
A_1\\
A_2\\
A_3 
\end{pmatrix}\,,
\end{align}
and
\begin{align}
\begin{pmatrix}
A_1\\
A_2\\
A_3
\end{pmatrix}=\frac{1}{(d-4)}  \begin{pmatrix}
d-2 & -1 & -1\\
-1 & d-2 & -1\\
-1 & -1 & d-2
\end{pmatrix} \begin{pmatrix}
\tilde{A}_1\\
\tilde{A}_2\\
\tilde{A}_3 
\end{pmatrix}\,.
\end{align}
In the above, the right hand side appears to introduce the spurious pole $1/(d-4)=-1/(2\epsilon)$. On the other hand, the sum
\begin{eqnarray}
A_{S}= A_1+A_2+A_3&=&\tilde{A}_1+\tilde{A}_2+\tilde{A}_3
\end{eqnarray}
is free of the spurious pole. As we discuss later, the amplitude only depends on $A_{S}$. Hence, there are no spurious poles.

It is straightforward to define the projectors for the coefficients $B^1_{kl}$ via
\begin{eqnarray}
\tilde{B}^1_{kl}&=&\underbrace{\frac{1}{(d-3)}\mathcal{P}_{\mu_1\mu_2}\mathcal{R}_{k,\mu_3}\mathcal{R}_{l,\mu_4}\mathcal{M}^{\mu_1\mu_2\mu_3\mu_4}}_{= \hat{B}^1_{kl}}-\left[\mathbb{G}^{-1}\right]_{kl}A_1\nonumber\\
&=&\frac{1}{(d-3)}\sum_{j_1,j_2=1}^{3}{\left[(d-3)B^1_{j_1j_2}\delta_{kj_1}\delta_{lj_2}\right]}=B^1_{kl}.
\end{eqnarray}
Similarly, we have
\begin{eqnarray}
B^2_{kl}&=&\frac{1}{(d-3)}\mathcal{P}_{\mu_1\mu_3}\mathcal{R}_{k,\mu_2}\mathcal{R}_{l,\mu_4}\mathcal{M}^{\mu_1\mu_2\mu_3\mu_4}-\left[\mathbb{G}^{-1}\right]_{kl}A_2,\nonumber\\
B^3_{kl}&=&\frac{1}{(d-3)}\mathcal{P}_{\mu_1\mu_4}\mathcal{R}_{k,\mu_2}\mathcal{R}_{l,\mu_3}\mathcal{M}^{\mu_1\mu_2\mu_3\mu_4}-\left[\mathbb{G}^{-1}\right]_{kl}A_3,\nonumber\\
B^4_{kl}&=&\frac{1}{(d-3)}\mathcal{P}_{\mu_2\mu_3}\mathcal{R}_{k,\mu_1}\mathcal{R}_{l,\mu_4}\mathcal{M}^{\mu_1\mu_2\mu_3\mu_4}-\left[\mathbb{G}^{-1}\right]_{kl}A_3,\nonumber\\
B^5_{kl}&=&\frac{1}{(d-3)}\mathcal{P}_{\mu_2\mu_4}\mathcal{R}_{k,\mu_1}\mathcal{R}_{l,\mu_3}\mathcal{M}^{\mu_1\mu_2\mu_3\mu_4}-\left[\mathbb{G}^{-1}\right]_{kl}A_2,\nonumber\\
B^6_{kl}&=&\frac{1}{(d-3)}\mathcal{P}_{\mu_3\mu_4}\mathcal{R}_{k,\mu_1}\mathcal{R}_{l,\mu_2}\mathcal{M}^{\mu_1\mu_2\mu_3\mu_4}-\left[\mathbb{G}^{-1}\right]_{kl}A_1\,.
\end{eqnarray}
Using Bose symmetry and Ward identities, we can derive the following relations between the form factors
\begin{eqnarray}
\Delta B^1_{11}(s,t,u)&=&B^1_{11}(s,t,u)-B^1_{12}(s,t,u)=\underbrace{\left(\hat{B}^1_{11}(s,t,u)-\hat{B}^1_{12}(s,t,u)\right)}_{=\Delta \hat{B}^1_{11}(s,t,u)}-\frac{A_1(s,t,u)}{u}.\nonumber\\
\end{eqnarray}

With the same spirit, we can define the projectors for the coefficients $C_{ijkl}$ via
\begin{eqnarray}
\tilde{C}_{ijkl}&=&\underbrace{\mathcal{R}_{i,\mu_1}\mathcal{R}_{j,\mu_2}\mathcal{R}_{k,\mu_3}\mathcal{R}_{l,\mu_4}\mathcal{M}^{\mu_1\mu_2\mu_3\mu_4}}_{=\hat{C}_{ijkl}}-\left\{\left[\mathbb{G}^{-1}\right]_{ij}B^1_{kl}+
\left[\mathbb{G}^{-1}\right]_{ik}B^2_{jl}+\left[\mathbb{G}^{-1}\right]_{il}B^3_{jk}\right.\nonumber\\
&&\left.+\left[\mathbb{G}^{-1}\right]_{jk}B^4_{il}+\left[\mathbb{G}^{-1}\right]_{jl}B^5_{ik}
+\left[\mathbb{G}^{-1}\right]_{kl}B^6_{ij}\right\}\nonumber\\
&&-\left\{\left[\mathbb{G}^{-1}\right]_{ij}
\left[\mathbb{G}^{-1}\right]_{kl}A_1+\left[\mathbb{G}^{-1}\right]_{ik}
\left[\mathbb{G}^{-1}\right]_{jl}A_2+\left[\mathbb{G}^{-1}\right]_{il}
\left[\mathbb{G}^{-1}\right]_{jk}A_3\right\}\nonumber\\
&=&C_{ijkl}\,.
\end{eqnarray}
Furthermore, by using the relations between form factors from Bose symmetry and Ward identities, we obtain
\begin{align}
\Delta C_{2111}(s,t,u)&=C_{2111}(s,t,u)-C_{2112}(s,t,u) \nonumber\\ &=\underbrace{\left(\hat{C}_{2111}(s,t,u)-\hat{C}_{2112}(s,t,u)\right)}_{=\Delta \hat{C}_{2111}(s,t,u)}+\; \frac{A_S(s,t,u)}{su}.\nonumber\\
\end{align}

With the above preparations, the helicity-summed amplitude square takes the form
\begin{eqnarray}
&&\sum_{\lambda_1,\lambda_2,\lambda_3,\lambda_4=\pm}{\left|\mathcal{M}_{\lambda_1\lambda_2\lambda_3\lambda_4}\right|^2}=2\left|A_S(s,t,u)\right|^2+2\sum_{i=1}^{3}{\left|A_i(s,t,u)\right|^2}\nonumber\\
&&\qquad -2su\Re{\left(\Delta C_{2111}(s,t,u) A_S^*(s,t,u)\right)}+s^2u^2\left|\Delta C_{2111}(s,t,u)\right|^2\nonumber\\
&&\qquad +4\left[u\Re{\left(\Delta B^1_{11}(s,t,u)A_1^*(s,t,u)\right)}+s\Re{\left(\Delta B^1_{11}(t,u,s)A_1^*(t,u,s)\right)}+t\Re{\left(\Delta B^1_{11}(u,s,t)A_1^*(u,s,t)\right)}\right]\nonumber\\
&&\qquad +2u^2\left|\Delta B^1_{11}(s,t,u)\right|^2+2s^2\left|\Delta B^1_{11}(t,u,s)\right|^2+2t^2\left|\Delta B^1_{11}(u,s,t)\right|^2\label{eq:helsuminFF1}\\
&&\quad =\left|A_S(s,t,u)\right|^2+2u^2\left|\Delta \hat{B}^1_{11}(s,t,u)\right|^2+2s^2\left|\Delta \hat{B}^1_{11}(t,u,s)\right|^2+2t^2\left|\Delta \hat{B}^1_{11}(u,s,t)\right|^2 \label{eq:helsuminFF2}\\
&&\qquad +s^2u^2\left|\Delta \hat{C}_{2111}(s,t,u)\right|^2,\nonumber
\end{eqnarray}
where we have used the relations
\begin{eqnarray}
s\Delta B^1_{11}(t,u,s)&=&u\Delta B^1_{11}(t,s,u), \quad s\Delta \hat{B}^1_{11}(t,u,s)=u\Delta \hat{B}^1_{11}(t,s,u),\quad A_1(t,u,s)=A_1(t,s,u),\nonumber\\
A_2(s,t,u)&=&A_1(u,s,t),\quad A_3(s,t,u)=A_1(t,s,u).
\end{eqnarray}
As we see from eq.\eqref{eq:helsuminFF2}, at the level of helicity-summed amplitude square, we only need to calculate $5$ independent form factors $A_S(s,t,u)$, $\Delta \hat{B}^1_{11}(s,t,u), \;\Delta \hat{B}^1_{11}(u,s,t), \;\Delta \hat{B}^1_{11}(t,u,s)$, and $\Delta \hat{C}_{2111}(s,t,u)$. Among these, there are no interference terms. The form factors $A_S(s,t,u),su\Delta C_{2111}(s,t,u),$ and $su\Delta \hat{C}_{2111}(s,t,u)$ are fully symmetric in the Mandelstam
variables $s,t,u$.


Furthermore, the helicity amplitudes can be expressed as a linear combination of the form factors. Like the case of form factors, there are $5$ independent helicity amplitudes.\footnote{Note that our convention of helicities is different with respect to ref.~\cite{Bern:2001dg}, because we take all momenta as incoming. The helicities of the final particles in ref.~\cite{Bern:2001dg} should be flipped in order to match to our case.} Up to a global phase, we can write them as:
\begin{eqnarray}
\mathcal{M}_{++++}&=&A_S(s,t,u)+\frac{u}{2}\Delta B^1_{11}(s,t,u)+\frac{s}{2}\Delta B^1_{11}(t,u,s)
+\frac{t}{2}\Delta B^1_{11}(u,s,t)-\frac{su}{4}\Delta C_{2111}(s,t,u)\nonumber\\
&=&\frac{1}{4}A_S(s,t,u)+\frac{u}{2}\Delta \hat{B}^1_{11}(s,t,u)+\frac{s}{2}\Delta \hat{B}^1_{11}(t,u,s)+\frac{t}{2}\Delta \hat{B}^1_{11}(u,s,t)-\frac{su}{4}\Delta \hat{C}_{2111}(s,t,u),\nonumber\\ \\
\mathcal{M}_{-+++}&=&\frac{su}{4}\Delta C_{2111}(s,t,u)=\frac{1}{4}A_S(s,t,u)+\frac{su}{4}\Delta \hat{C}_{2111}(s,t,u),\\
\mathcal{M}_{--++}&=&A_1(s,t,u)+\frac{u}{2}\Delta B^1_{11}(s,t,u)-\frac{s}{2}\Delta B^1_{11}(t,u,s)-\frac{t}{2}\Delta B^1_{11}(u,s,t)-\frac{su}{4}\Delta C_{2111}(s,t,u)\nonumber\\
&=&\frac{1}{4}A_S(s,t,u)+\frac{u}{2}\Delta \hat{B}^1_{11}(s,t,u)-\frac{s}{2}\Delta \hat{B}^1_{11}(t,u,s)-\frac{t}{2}\Delta \hat{B}^1_{11}(u,s,t)-\frac{su}{4}\Delta \hat{C}_{2111}(s,t,u),\nonumber\\ \\
\mathcal{M}_{+-+-}&=&A_1(u,s,t)-\frac{u}{2}\Delta B^1_{11}(s,t,u)-\frac{s}{2}\Delta B^1_{11}(t,u,s)+\frac{t}{2}\Delta B^1_{11}(u,s,t)-\frac{su}{4}\Delta C_{2111}(s,t,u)\nonumber\\
&=&\frac{1}{4}A_S(s,t,u)-\frac{u}{2}\Delta \hat{B}^1_{11}(s,t,u)-\frac{s}{2}\Delta \hat{B}^1_{11}(t,u,s)+\frac{t}{2}\Delta \hat{B}^1_{11}(u,s,t)-\frac{su}{4}\Delta \hat{C}_{2111}(s,t,u),\nonumber\\ \\
\mathcal{M}_{+--+}&=&A_1(t,s,u)-\frac{u}{2}\Delta B^1_{11}(s,t,u)+\frac{s}{2}\Delta B^1_{11}(t,u,s)-\frac{t}{2}\Delta B^1_{11}(u,s,t)-\frac{su}{4}\Delta C_{2111}(s,t,u)\nonumber\\
&=&\frac{1}{4}A_S(s,t,u)-\frac{u}{2}\Delta \hat{B}^1_{11}(s,t,u)+\frac{s}{2}\Delta \hat{B}^1_{11}(t,u,s)-\frac{t}{2}\Delta \hat{B}^1_{11}(u,s,t)-\frac{su}{4}\Delta \hat{C}_{2111}(s,t,u).\nonumber\\\label{eq:ampinff}
\end{eqnarray}
In addition, if we take into account permutations, the following relations hold between the $3$ independent helicity amplitudes:
\begin{eqnarray}
\mathcal{M}_{+-+-}&=&\left.\mathcal{M}_{--++}\right|_{s\leftrightarrow u}=\left.\mathcal{M}_{+--+}\right|_{t\leftrightarrow u}\label{eq:Mmmpprelation} \,.
\end{eqnarray}
While the helicity amplitudes $\mathcal{M}_{++++}$ and $\mathcal{M}_{-+++}$ are fully symmetric in the arguments $(s,t,u)$, 
the remaining ones $\{\mathcal{M}_{--++},~\mathcal{M}_{+-+-},~\mathcal{M}_{+--+}\} $ are symmetric in $\{(t,u),(s,t),(s,u)\}$ respectively.
The amplitudes in other helicity configurations can be related to the above mentioned $5$ as follows:
\begin{eqnarray}
\mathcal{M}_{----}&=&\mathcal{M}_{++++},\quad \mathcal{M}_{++--}=\mathcal{M}_{--++},\quad \mathcal{M}_{-+-+}=\mathcal{M}_{+-+-},\quad \mathcal{M}_{-++-}=\mathcal{M}_{+--+},\nonumber\\
\mathcal{M}_{+-++}&=&\mathcal{M}_{++-+}=\mathcal{M}_{+++-}=\mathcal{M}_{+---}=\mathcal{M}_{-+--}=\mathcal{M}_{--+-}=\mathcal{M}_{---+}=\mathcal{M}_{-+++}.\nonumber\\
\end{eqnarray}
Summing these helicity amplitudes, it is straightforward to verify that eqs.\eqref{eq:helsuminFF1} and \eqref{eq:helsuminFF2} hold. 

\section{Two-loop master integrals}
\label{sec:masterintegrals2L}

It is well known that a form factor or an amplitude at $L$-loop can be written as a linear combination of a small set of $L$-loop Feynman integrals, known as master integrals.
In this section, we first present the details for obtaining analytic expressions of two-loop master integrals for LbL scattering, before discussing in detail their numerical evaluations. The one-loop master integrals are provided in appendix \ref{sec:masterintegrals1L}. In dimensional regularisation, the two-loop QED amplitude for a given fermion species with mass $m_f$, generated by \qgraf~\cite{Nogueira:1991ex} and \feynarts~\cite{Hahn:2000kx}, can be reduced into 103 $d$-dimensional master integrals across all the permutations by applying integration-by-parts (IBP) identities with the help of \kira~\cite{Maierhofer:2017gsa,Klappert:2020nbg} or \finiteflow~\cite{Peraro:2019svx}+\litered~\cite{Lee:2012cn}. Modulo the permutations in the Mandelstam variables $s,t,u$, there are $30$ master integrals. One among them is a product of a one-loop box integral and a one-loop tadpole integral, which can be solved separately. The remaining $29$ integrals form a closed system in the method of differential equations~\cite{Kotikov:1990kg,Remiddi:1997ny,Gehrmann:1999as,Argeri:2007up} in $d$ dimensions. We now explain how to analytically solve these 29 master integrals.

The two-loop master integrals for LbL in the top-sector have been investigated in ref.~\cite{Caron-Huot:2014lda}. In this context, the new ingredients of this article is the presentation of the analytic results in a form that is most suitable for phenomenological applications, facilitating the fast numerical evaluation of the scattering amplitudes across all the physical phase-space regions. In addition, to ensure the comprehensiveness of this article, we elaborate on all the essential steps needed to replicate our results, even at the risk of providing details that may already be well understood. We express the analytic results of the master integrals in terms of Chen iterated integrals~\cite{Chen:1977oja} with logarithmic one-forms, and cast the results in a form suitable for phenomenological applications. 
The family of master integrals of interest is expressed as
\begin{align}
\label{eq:define-int-2L}
I_{a_1,\cdots,a_9}^{(2)}(s_{ij},s_{jk},s_{ik}) &= \left(\frac{e^{\epsilon \gamma_E}m_f^{2\epsilon}}{i\pi^{\frac{d}{2}}}\right)^2 \int \derive^d\ell_1 \derive^d\ell_2 \frac{1}{D_1^{a_1} D_2^{a_2} D_3^{a_3}D_4^{a_4}D_5^{a_5}D_6^{a_6}D_7^{a_7}D_8^{a_8} D_9^{a_9}}\,,
\end{align}
where $a_i \in \mathbb{Z}$. We represent the loop momenta through $\ell_i$ and Euler-Mascheroni constant by $\gamma_E$. The inverse propagators are
\begin{eqnarray}
D_1&=&\ell_1^2-m_f^2+i0^+,\quad D_2=(\ell_1+p_i)^2-m_f^2+i0^+,\quad D_3=(\ell_1+p_i+p_j)^2-m_f^2+i0^+,\nonumber\\
D_4&=&(\ell_1+p_i+p_j+p_k)^2-m_f^2+i0^+,\qquad D_5=\ell_2^2-m_f^2+i0^+,\nonumber \\  D_6&=&(\ell_2+p_i)^2-m_f^2+i0^+,\qquad 
D_7=(\ell_2+p_i+p_j)^2-m_f^2+i0^+,\nonumber \\  D_8&=&(\ell_2+p_i+p_j+p_k)^2-m_f^2+i0^+,\qquad D_9=(\ell_1-\ell_2)^2+i0^+.\nonumber\\
\end{eqnarray}
The introduction of $i0^+$ ensures the prescription of the Feynman propagators. We keep the general external momentum dependence with $i,j,k\in\ \left\{1,2,3,4\right\}$ and $i\neq j, i\neq k, j\neq k$. The arguments of $I_{a_1,\cdots,a_9}^{(2)}$ are defined as $s_{ij}=(p_i+p_j)^2,s_{jk}=(p_j+p_k)^2,s_{ik}=(p_i+p_k)^2$, which are simply a permutation of $s,t,u$.

As mentioned above, to obtain the analytic results of the master integrals we use the method of differential equations. Within this approach, it is convenient to choose a canonical basis of master integrals that allow us to set up the differential equations into the so-called $\epsilon$-form~\cite{Henn:2013pwa}, which enables to solve the system afterwards.  Our choice of the canonical basis for the two-loop master integrals is guided by the one outlined in ref.~\cite{Caron-Huot:2014lda}, although not completely identical. For completeness of this article, we present our chosen basis  $\vec{f}^{(2)}=(f^{(2)}_1,\cdots,f^{(2)}_{29})$ in eq.\eqref{eq:UTbasis4mf} in appendix \ref{sec:canonmasterintegrals2L}.
Each integral in the canonical basis eq.\eqref{eq:UTbasis4mf} is dimensionless and has uniform transcendental (UT) weight properties. Moreover, the system of differential equations of $\vec{f}^{(2)}$ attains the form
\begin{equation}
    \derive \vec{f}^{(2)}= \epsilon\; \derive A^{(2)} \vec{f}^{(2)},\label{eq:DE1}
\end{equation}
where the differential equation matrix $A^{(2)}$ does not depend on $\epsilon$. Due to the inclusion of massive loop propagators, the presence of square roots is unavoidable. The square roots present in our case are $\sqrt{s_{ij}(s_{ij}-4m_f^2)}$,$\sqrt{s_{jk}(s_{jk}-4m_f^2)}$, $\sqrt{s_{ij} s_{jk} (s_{ij}s_{jk} -4m_f^2(s_{ij}+s_{jk}))}$ and $\sqrt{s_{ij} (m_f^4 s_{ij}-2m_f^2 s_{jk} (s_{ij}+2 s_{jk})+s_{ij} s_{jk}^2)}$. We can rationalise the first three square roots simultaneously by choosing the same variables as suggested in ref.~\cite{Caron-Huot:2014lda}:
\begin{eqnarray}
\label{eq:parametrisation}
\frac{s_{ij}}{m_f^2}&=&-\frac{4(w-z)^2}{(1-w^2)(1-z^2)},\quad \frac{s_{jk}}{m_f^2}=-\frac{(w-z)^2}{wz},\quad s_{ik}=-s_{ij}-s_{jk}.
\end{eqnarray}
However, we still have one square root $\rho=\sqrt{-2 w z + z^2 + w^4 z^2 - 2 w^3 z^3 + w^2 (1 + z^2 + z^4)}$ left, which we are unable to rationalise further. There are in general multiple (complex) solutions of eq.\eqref{eq:parametrisation} for $w,z$ in terms of $s_{ij},s_{jk}$. Our choice of the solutions, depending on the real values of $s_{ij}$ and $s_{jk}$, is given in appendix \ref{sec:wzdef}.

The next step is to integrate the system of differential equations to obtain the analytic results. Our goal is to express the analytic results in terms of iterated integrals with polylogarithmic kernels, with explicit representation in terms of Goncharov polylogarithms (GPLs)~\cite{goncharov2011multiple} [\ie, multiple polylogarithms (MPLs)] wherever possible. We start with rewriting the one-forms as logarithmic one-forms and casting the differential equations in a more compact form. Feynman integrals expressed in terms of iterated integrals can be efficiently described via their \textit{symbols}~\cite{Goncharov:2010jf}. In the variables defined in eq.~\eqref{eq:parametrisation}, the \textit{alphabet} $\mathbb{A}$ consists of the following $16$ \textit{letters}:
\begin{align}
\mathbb{A}=&\left\{l_1(w,z),\cdots,l_{16}(w,z)\right\}\nonumber\\
=&\left\{1 - w, 1 + w, 1 - w z, w - z, w, w + z, 1 + w - z + w z, 1 - w + z +  w z, \right.\nonumber\\ & 1 + w z, 1 - z, 1 + z,  z,
 \frac{-\rho + w - z - w z + w^2 z - w z^2}{\rho + w - z - w z + w^2 z - w z^2}, \frac{-\rho + w^2 - 3 w z + z^2}{\rho + w^2 - 3 w z + z^2},  \nonumber \\ &\left.\frac{-1 - \rho + w^2 z - w z^2}{-1 + \rho + w^2 z - w z^2}, \frac{1 - \rho + w^2 z - w z^2}{1 + \rho + w^2 z - w z^2}\right\},
\end{align}
where the last four depend on the square root $\rho$. In other words, the matrix $A^{(2)}$ in the differential equation \eqref{eq:DE1} can be expressed in terms of these $16$ one-forms:
\begin{equation}
    \derive A_{i,j}^{(2)}=\sum_{k=1}^{16}{c_{ijk}\dlog(l_k(w,z))}, \qquad  c_{ijk}\in \mathbb{Q}.
\end{equation}
Furthermore, this matrix can be rewritten in terms of the following $15$ independent linear combinations of $\dlog$ one-forms:
\begin{eqnarray}
\derive g_1&=&-\dlog(l_1)+\dlog(l_2)+\dlog(l_{10})-\dlog(l_{11}),\nonumber\\
\derive g_2&=&\dlog(l_1)+\dlog(l_2)-2\dlog(l_3)+\dlog(l_{10})+\dlog(l_{11}),\nonumber\\
\derive g_3&=&-\dlog(l_1)-\dlog(l_2)+2\dlog(l_4)-\dlog(l_{10})-\dlog(l_{11}),\nonumber\\
\derive g_4&=&2\dlog(l_4)-\dlog(l_5)-\dlog(l_{12}),\nonumber\\
\derive g_5&=&-\dlog(l_5)+\dlog(l_{12}),\nonumber\\
\derive g_6&=&-2\dlog(l_4)+4\dlog(l_5) -6\dlog(l_6)+4\dlog(l_{12}),\nonumber\\
\derive g_7&=&\frac{1}{2}\dlog(l_{14}), \nonumber\\
\derive g_8&=&-\dlog(l_{15})+\dlog(l_{16}),\nonumber\\
\derive g_9&=&-2\dlog(l_{13})+\dlog(l_{14})+\dlog(l_{15})+\dlog(l_{16}),\nonumber\\
\derive g_{10}&=&\dlog(l_4)-\dlog(l_5)+\frac{1}{2}\dlog(l_7)+\frac{1}{2}\dlog(l_8)-\dlog(l_{12}),\nonumber\\
\derive g_{11}&=&2\dlog(l_{13})-\dlog(l_{14}), \nonumber\\
\derive g_{12}&=&-\dlog(l_5)-\dlog(l_{12}),\nonumber\\
\derive g_{13}&=&-\dlog(l_7)+\dlog(l_8),\nonumber\\
\derive g_{14}&=&\frac{1}{2}\dlog(l_1)-\frac{1}{2}\dlog(l_2)+\frac{1}{2}\dlog(l_{10})-\frac{1}{2}\dlog(l_{11}),\nonumber\\ 
\derive g_{15}&=&-2\dlog(l_{4})+2\dlog(l_5)+2\dlog(l_7)+2\dlog(l_8)-6\dlog(l_9)+2\dlog(l_{12}).\label{eq:oneforms}
\end{eqnarray}
Here we have dropped the arguments $(w,z)$ in the one-forms. The non-zero matrix elements of $A^{(2)}$ are given in eq.\eqref{eq:A2ME} in appendix \ref{sec:canonmasterintegrals2L}. In the next step, we take the series expansion over the dimensional regulator $\epsilon$ in the master integrals $\vec{f}^{(2)}=\sum_{\omega=0}{\epsilon^\omega \vec{f}^{(2,\omega)}}$ and the differential equation system eq.\eqref{eq:DE1} becomes
\begin{eqnarray}
\derive \vec{f}^{(2,\omega+1)}&=&\derive A^{(2)}\vec{f}^{(2,\omega)},\quad \omega\in \mathbb{N}.\label{eq:DE2}
\end{eqnarray}
Since our choice of $\vec{f}^{(2)}$ is a UT basis, all terms contributing to $\vec{f}^{(2,\omega)}$ should have the same transcendental weight $\omega$. With a proper boundary condition, we can solve eq.\eqref{eq:DE2} iteratively from lower to higher weights. The solution can be written in terms of iterated integrals with logarithmic one-forms. Up to weight $\omega=4$ that we need, most of the master integrals $f_n^{(2,\omega)}(x_{ij},x_{jk},x_{ik})$ can be expressed through Goncharov polylogarithm $G$'s, while only the iterated integrals for the following master integrals depend on the kernels with the square root $\rho$:
\begin{equation}
f_{18}^{(2,3)},f_{18}^{(2,4)},f_{19}^{(2,4)},f_{20}^{(2,4)},f_{25}^{(2,4)},f_{26}^{(2,4)},f_{28}^{(2,4)},f_{29}^{(2,4)}.
\end{equation}
In our two-loop amplitudes, we actually only need 5 integrals $f_{18}^{(2,3)},f_{25}^{(2,4)},f_{26}^{(2,4)},f_{28}^{(2,4)}$, and $f_{29}^{(2,4)}$ which cannot be written in $G$'s. For the integrals with $G$'s,~\footnote{Their concrete expressions are available in the ancillary files.} their numerical evaluations are rather straightforward with the algorithm outlined in ref.~\cite{Vollinga:2004sn}. In this paper, we use \fastgpl~\cite{Wang:2021imw} and take \handyg~\cite{Naterop:2019xaf} as a cross check.

Now, as a new ingredient from this section, we outline the steps taken to prepare all the master integrals present in the scattering amplitudes for phenomenological applications. We explain how to numerically evaluate iterated integrals with algebraic kernels using an idea that has been recently successfully utilised for other processes (see, \eg, ref.~\cite{Chicherin:2020oor}). We also find different analytic expressions for all the master integrals valid in different regions of phase space in order to avoid performing analytic continuations. Moreover, we identify relations among the Feynman integrals in the scattering amplitudes, which reduces the number of iterated integrals to be computed numerically at each point of the phase space. For the numerical evaluations of our master integrals with square roots, we choose a technique elucidated in refs.~\cite{Caron-Huot:2014lda,Chicherin:2020oor}. The kernels in the first two-fold integration can be rationalised and therefore can be expressed through logarithms and classical polylogarithms $\mathrm{Li}_n$ in terms of $x_{ij},x_{jk},x_{ik}$.~\footnote{We choose the classical polylogarithms rather than the GPLs here because the numerical evaluations of the former are much faster than the latter. This is possible thanks to the methods outlined in ref.~\cite{Duhr:2011zq} by matching symbols in both the $(w,z)$ as well as $(s_{ij},s_{jk})$ coordinate systems. However, the Riemann sheets for the multi-valued functions should be carefully picked up.} For the weight-$3$ integral $f_{18}^{(2,3)}$, we are only left with a single integration to be carried out numerically, while there are two-fold integrations left for the weight-$4$ integrals. For the latter, the basic idea is to convert the last two-fold integration into a one-fold integration, and to perform the last one-fold integration numerically (cf. an example given in section 4.4 of ref.~\cite{Caron-Huot:2014lda}). For numerical integration in all phase-space regions in appendix~\ref{sec:wzdef}, we also use an approach that makes the numerical evaluation of all the master integrals faster as we avoid crossing any physical singularity. Instead of integrating our results within a single region and then relying on analytic continuation to cross the boundaries into other regions, we obtain distinct analytic results that are specifically valid within each of these regions. Hence, our numerical evaluations of the integrals are valid within specific phase-space regions. We also obtain analytic boundary constants $f^{(2,\omega)}_{n}(x_{ij},0,-x_{ij})$ and $f^{(2,\omega)}_{n}(0,x_{jk},-x_{jk})$ valid in all these regions. For these analytic boundary constants for the $5$ iterated integrals with roots, refer to appendix \ref{app:BCs}. For the last one-fold integration, the convergence of the numerical integration strongly depends on the path. We opt for the following path parameterisation from the starting point $(s_{ij,0},s_{jk,0})$ to the end point $(s_{ij,1},s_{jk,1})$
\begin{eqnarray}
s_{ij}(r)&=&s_{ij,0}(1-r)^5\left(1+5r+15r^2+35r^3+70r^4\right)\nonumber\\
&&+s_{ij,1}r^5\left(126-420r+540r^2-315r^3+70r^4\right),\nonumber\\
s_{jk}(r)&=&s_{jk,0}(1-r)^5\left(1+5r+15r^2+35r^3+70r^4\right)\nonumber\\
&&+s_{jk,1}r^5\left(126-420r+540r^2-315r^3+70r^4\right),\label{eq:path4int}
\end{eqnarray}
where $r\in [0,1]$. It is easy to verify that $(s_{ij}(0),s_{jk}(0))=(s_{ij,0},s_{jk,0})$ and $(s_{ij}(1),s_{jk}(1))=(s_{ij,1},s_{jk,1})$. In this article, we choose the following boundary conditions:
\begin{eqnarray}
\left(s_{ij,0},s_{jk,0}\right)&=&\left\{\begin{array}{ll}(0,0), & \quad {\rm if}~s_{ij},s_{jk}<0\\
(s_{ij},0), & \quad {\rm if}~s_{ij}>0,s_{jk}<0\\
(0,s_{jk}), & \quad {\rm if}~s_{ij}<0,s_{jk}>0.\end{array}\right.\,
\end{eqnarray}
The integrals have been cross checked numerically for a few phase space points in each region against \amflow~\cite{Liu:2022chg}.

As the final step in making the integrals pheno-ready and in order to simplify the final helicity amplitudes, it is necessary to find all possible relations among the master integrals needed for our amplitudes. Otherwise, not only can we not check pole cancellations at the analytic level, as a similar finding in ref.~\cite{Badger:2021owl} for another process, but large numerical cancellations will also happen among different pieces, which prevents us from taking any interesting kinematic limit. In this work, we use the symbol techniques~\cite{Duhr:2019tlz} and the shuffle algebra that the iterated integrals satisfy to find the following relations, by including the one-loop master integrals $f_n^{(1,\omega)}$ defined in appendix~\ref{sec:masterintegrals1L}. The relations can be categorised into three parts:
\begin{itemize}
    \item The following integrals are identical over permutations of external momenta:
    \begin{align}
    f_n^{(1,\omega)}(x_{ij},x_{jk},x_{ik}) =&~ f_n^{(1,\omega)}(x_{ik},x_{jk},x_{ij}) \quad \text{for}~ n\in\{3,5\}\,,
    \nonumber \\
    f_n^{(1,\omega)}(x_{ij},x_{jk},x_{ik}) =&~f_n^{(1,\omega)}(x_{ij},x_{ik},x_{jk}) \quad \text{for}~ n\in\{2,4\}\,,
    \nonumber \\
    f_6^{(1,\omega)}(x_{ij},x_{jk},x_{ik}) =&~ f_6^{(1,\omega)}(x_{jk},x_{ij},x_{ik})\,,
    \nonumber \\
    f_n^{(2,\omega)}(x_{ij},x_{jk},x_{ik}) =&~ f_n^{(2,\omega)}(x_{ij},x_{ik},x_{jk}) \quad \text{for}~ n\in\{2,3,4,5,6,7,8,9,10,11,12,13,24\} \,,\nonumber \\
    f_n^{(2,\omega)}(x_{ij},x_{jk},x_{ik}) =&~ f_n^{(2,\omega)}(x_{ik},x_{jk},x_{ij}) \quad \text{for}~ n\in\{14,15,16,17\}\,,\nonumber \\
    f_{19}^{(2,2)}(x_{ij},x_{jk},x_{ik}) =&~ f_{19}^{(2,2)}(x_{jk},x_{ij},x_{ik})\,,\nonumber\\
    f_{n}^{(2,\omega)}(x_{ij},x_{jk},x_{ik}) =&~ f_{n}^{(2,\omega)}(x_{jk},x_{ij},x_{ik}) \quad \text{for}~ n\in\{21,22,23\}\,. \label{eq:fnrelation1}
    \end{align}
    \item Some integrals with same transcendental weight can be related:
    \begin{align}
    f_2^{(1,\omega)}(x_{ij},x_{jk},x_{ik}) =&~ f_{3}^{(1,\omega)}(x_{jk},x_{ij},x_{ik}),\quad
    f_4^{(1,\omega)}(x_{ij},x_{jk},x_{ik}) = ~f_{5}^{(1,\omega)}(x_{jk},x_{ij},x_{ik})\,,\nonumber \\
    f_{6}^{(2,\omega)}(x_{ij},x_{jk},x_{ik}) =&~ f_{14}^{(2,\omega)}(x_{jk},x_{ij},x_{ik}),\quad
    f_{7}^{(2,\omega)}(x_{ij},x_{jk},x_{ik}) =~ f_{15}^{(2,\omega)}(x_{jk},x_{ij},x_{ik}) \,,\nonumber \\
    f_{8}^{(2,\omega)}(x_{ij},x_{jk},x_{ik}) =&~ f_{16}^{(2,\omega)}(x_{jk},x_{ij},x_{ik}),\quad f_{9}^{(2,2)}(x_{ij},x_{jk},x_{ik})=f_{16}^{(2,2)}(x_{jk},x_{ij},x_{ik})\,,\nonumber\\
    f_{15}^{(2,1)}(x_{ij},x_{jk},x_{ik}) =& -2 f_2^{(2,1)}(x_{jk},x_{ij},x_{ik}),\quad
    f_3^{(2,2)}(x_{ij},x_{jk},x_{ik})= 4 f_{16}^{(2,2)}(x_{jk},x_{ij},x_{ik})\,,\nonumber \\
    f_4^{(2,2)}(x_{ij},x_{jk},x_{ik}) =&~ 2 f_{16}^{(2,2)}(x_{jk},x_{ij},x_{ik}),\quad
    f_{14}^{(2,2)}(x_{ij},x_{jk},x_{ik}) =-4f_{16}^{(2,2)}(x_{ij},x_{jk},x_{ik})\,,\nonumber \\
    f_{22}^{(2,2)}(x_{ij},x_{jk},x_{ik}) =&~ 2 f_{19}^{(2,2)}(x_{ij},x_{jk},x_{ik}),\quad f_4^{(2,3)}(x_{ij},x_{jk},x_{ik})= -f_5^{(1,3)}(x_{jk},x_{ij},x_{ik})\,,\nonumber \\
    f_2^{(2,\omega)}(x_{ij},x_{jk},x_{ik}) =& -f_2^{(1,\omega)}(x_{ij},x_{jk},x_{ik}) \qquad \text{for}~ \omega\in\{1,2\}\,,\nonumber\\
    f_4^{(2,\omega)}(x_{ij},x_{jk},x_{ik}) =& -f_4^{(1,\omega)}(x_{ij},x_{jk},x_{ik}) \qquad \text{for}~ \omega\in\{2,3\}\,,\nonumber\\
    f_{11}^{(2,2)}(x_{ij},x_{jk},x_{ik}) =& ~\frac{3}{2} f_3^{(1,2)}(x_{jk},x_{ij},x_{ik})-\frac{1}{4}f_{15}^{(2,2)}(x_{jk},x_{ij},x_{ik}) \,,\nonumber\\
    f_{8}^{(2,3)}(x_{ij},x_{jk},x_{ik}) =& -\frac{3}{4} f_{6}^{(2,3)}(x_{ij},x_{jk},x_{ik})+f_{10}^{(2,3)}(x_{ij},x_{jk},x_{ik})-3f_4^{(1,3)}(x_{ij},x_{jk},x_{ik}) \,,\nonumber\\
    f_{9}^{(2,3)}(x_{ij},x_{jk},x_{ik}) =&~ \frac{1}{4} f_{14}^{(2,3)}(x_{jk},x_{ij},x_{ik})-2f_5^{(1,3)}(x_{jk},x_{ij},x_{ik}) \,.\label{eq:fnrelation2}
    \end{align}
    \item Some higher transcendental weight two-loop master integrals can be expressed into lower weight integrals:
     \begin{align}
     f_5^{(1,2)}(x_{ij},x_{jk},x_{ik})=&~ \frac{1}{2}\left(f_3^{(1,1)}(x_{ij},x_{jk},x_{ik})\right)^2,\nonumber \\
     f_{24}^{(2,4)}(x_{ij},x_{jk},x_{ik}) =&~ \frac{1}{4}\left(f_{3}^{(1,1)}(x_{jk},x_{ij},x_{ik})\right)^4\, ,\nonumber\\
     f_{16}^{(2,2)}(x_{ij},x_{jk},x_{ik})=&-\frac{1}{4}\left(f_{3}^{(1,1)}(x_{ij},x_{jk},x_{ik})\right)^2\,,\nonumber \\
     f_{5}^{(2,3)}(x_{ij},x_{jk},x_{ik}) =& -\frac{1}{2}\left(f_{3}^{(1,1)}(x_{jk},x_{ij},x_{ik})\right)^3,\nonumber \\
     f_{3}^{(2,3)}(x_{ij},x_{jk},x_{ik})=&-2 f_{3}^{(1,1)}(x_{jk},x_{ij},x_{ik})f_{3}^{(1,2)}(x_{jk},x_{ij},x_{ik})\,,\nonumber \\
     f_{5}^{(2,4)}(x_{ij},x_{jk},x_{ik})=&-f_{2}^{(1,2)}(x_{ij},x_{jk},x_{ik})f_{4}^{(1,2)}(x_{ij},x_{jk},x_{ik})\;- \nonumber \\ & \hspace{4.5cm} f_{2}^{(1,1)}(x_{ij},x_{jk},x_{ik})f_{4}^{(1,3)}(x_{ij},x_{jk},x_{ik})\,,\nonumber \\
     f_{10}^{(2,3)}(x_{ij},x_{jk},x_{ik}) =&~\frac{1}{3}f_{14}^{(2,3)}(x_{jk},x_{ij},x_{ik})-2f_5^{(1,3)}(x_{jk},x_{ij},x_{ik})\;+\nonumber \\ 
     &\frac{1}{6} f_3^{(1,1)}(x_{jk},x_{ij},x_{ik}) \left(-6f_3^{(1,2)}(x_{jk},x_{ij},x_{ik}) + f_{15}^{(2,2)}(x_{jk},x_{ij},x_{ik})\right) \,,\nonumber \\
     f_{16}^{(2,3)}(x_{ij},x_{jk},x_{ik}) =&~ f_{5}^{(1,3)}(x_{ij},x_{jk},x_{ik})-\frac{5}{12}f_{14}^{(2,3)}(x_{ij},x_{jk},x_{ik})\;+\nonumber \\ 
     &\frac{1}{6} f_3^{(1,1)}(x_{ij},x_{jk},x_{ik}) \left(-6f_3^{(1,2)}(x_{ij},x_{jk},x_{ik}) + f_{15}^{(2,2)}(x_{ij},x_{jk},x_{ik})\right) \,,\nonumber \\
     f_{23}^{(2,3)}(x_{ij},x_{jk},x_{ik}) = &-\frac{1}{4}f_{10}^{(2,3)}(x_{ij},x_{jk},x_{ik})-\frac{1}{2}f_{10}^{(2,3)}(x_{jk},x_{ij},x_{ik})-f_{20}^{(2,3)}(x_{ij},x_{jk},x_{ik})\nonumber\\
     &+\frac{3}{2}f_{11}^{(2,2)}(x_{ij},x_{jk},x_{ik})f_{2}^{(2,1)}(x_{ij},x_{jk},x_{ik})\nonumber\\
     &+f_{11}^{(2,2)}(x_{jk},x_{ij},x_{ik})f_{2}^{(2,1)}(x_{jk},x_{ij},x_{ik})\,. \label{eq:fnrelation3}
    \end{align} 
\end{itemize}
We remind readers that $x_{ij}=s_{ij}/m_f^2,\;x_{jk}=s_{jk}/m_f^2,$ and $x_{ik}=s_{ik}/m_f^2$. Using the above relations, we manage to cancel the poles analytically. Furthermore, these relations also considerably reduce the number of master integrals that we need to compute at each transcendental weight from around 300 to 84 in the two-loop helicity amplitudes.

\section{Analytic expressions of helicity amplitudes}
\label{sec:analytichelamp}

In this section, we present our main results, \ie, the analytic expressions of helicity amplitudes for LbL. To obtain the helicity amplitudes we use \qgraf~ and \feynarts~ to generate the Feynman diagrams, and further manipulations such as performing $d$-dimensional Lorentz and Dirac algebra are done in {\tt FORM}~\cite{Vermaseren:2000nd,Ruijl:2017dtg} or {\tt Mathematica}.

We first present the one-loop form factors and helicity amplitudes as a warm-up exercise. In this section, we will only consider a given fermion loop with the fermion mass $m_f$, while the one-loop expressions for the $W^\pm$ loop can be found in appendix \ref{sec:oneloopamp4W}. Throughout this section, let us define the following dimensionless variables
\begin{equation}
x_s=\frac{s}{m_f^2},\quad x_t=\frac{t}{m_f^2},\quad x_u=\frac{u}{m_f^2}
\end{equation}
with $x_s+x_t+x_u=0$. Let us assume the fermion has charge $Q_f$ and colour charge $N_{c,f}$.~\footnote{In the SM, the quarks have $N_{c,f}=3$ and the leptons have $N_{c,f}=1$. The up-, down-type quarks have $Q_f=2/3$ and $Q_f=-1/3$, and the charged leptons have $Q_f=-1$.} In addition, we also denote the LO (one-loop) helicity amplitudes with the fermion $f$ as $\mathcal{M}_{\lambda_1\lambda_2\lambda_3\lambda_4}^{(0,0,f)}$, while the two-loop QCD and QED amplitudes with the fermion are $\mathcal{M}_{\lambda_1\lambda_2\lambda_3\lambda_4}^{(1,0,f)}$ and $\mathcal{M}_{\lambda_1\lambda_2\lambda_3\lambda_4}^{(0,1,f)}$ respectively.

\subsection{One-loop results}
\label{sec:oneloopamp}

The one-loop form factors are
\begin{eqnarray}
&&-iA_S(s,t,u)=4N_{c,f}Q_f^4\alpha^2\left\{2+\mathop{\sum_{(i,j,k)=(s,t,u),}}_{(t,u,s),(u,s,t)}{\left[r^{(1)}_1(x_i,x_j,x_k)\frac{f_2^{(1,1)}(x_i,x_j,x_k)}{\sqrt{x_i(x_i-4)}}\right.}\right.\nonumber\\
&&\qquad \qquad \qquad \left.\left.-r^{(1)}_2(x_i,x_j,x_k)f_4^{(1,2)}(x_i,x_j,x_k)-r^{(1)}_3(x_i,x_j,x_k)\frac{f_6^{(1,2)}(x_i,x_j,x_k)}{\sqrt{x_ix_j(x_ix_j+4x_k)}}\right]\right\},\nonumber\\
&&-i\Delta \hat{B}^1_{11}(s,t,u)=\frac{4N_{c,f}Q_f^4\alpha^2}{u}\left\{2-r^{(1)}_1(x_s,x_t,x_u)\frac{f_2^{(1,1)}(x_s,x_t,x_u)}{\sqrt{x_s(x_s-4)}}+r^{(1)}_4(x_t,x_u,x_s)\frac{f_2^{(1,1)}(x_t,x_u,x_s)}{\sqrt{x_t(x_t-4)}}\right.\nonumber\\
&& \qquad \qquad \qquad-r^{(1)}_4(x_u,x_s,x_t)\frac{f_2^{(1,1)}(x_u,x_s,x_t)}{\sqrt{x_u(x_u-4)}}+\left(r^{(1)}_5(x_s,x_t,x_u)+r^{(1)}_6(x_s,x_t,x_u)\right)f_4^{(1,2)}(x_s,x_t,x_u)\nonumber\\
&&\qquad \qquad \qquad+r^{(1)}_5(x_t,x_u,x_s)f_4^{(1,2)}(x_t,x_u,x_s)-r^{(1)}_5(x_u,x_s,x_t)f_4^{(1,2)}(x_u,x_s,x_t)\nonumber\\
&&\qquad \qquad \qquad-r^{(1)}_7(x_s,x_t,x_u)\frac{f_6^{(1,2)}(x_s,x_t,x_u)}{\sqrt{x_sx_t(x_sx_t+4x_u)}}+\left(r^{(1)}_7(x_t,x_u,x_s)-4x_u\right)\frac{f_6^{(1,2)}(x_t,x_u,x_s)}{\sqrt{x_tx_u(x_tx_u+4x_s)}}\nonumber\\
&&\qquad \qquad \qquad\left.-\left(r^{(1)}_7(x_u,x_s,x_t)+4(x_u-x_s)\right)\frac{f_6^{(1,2)}(x_u,x_s,x_t)}{\sqrt{x_ux_s(x_ux_s+4x_t)}}\right\},\nonumber\\
&&-i\Delta C_{2111}(s,t,u)=\frac{32N_{c,f}Q_f^4\alpha^2}{su}\left\{1+2\left(x_s^{-1}+x_t^{-1}+x_u^{-1}\right)\mathop{\sum_{(i,j,k)=(s,t,u),}}_{(t,u,s),(u,s,t)}{f_4^{(1,2)}(x_i,x_j,x_k)}\right.\nonumber\\
&&\qquad \qquad \qquad \left.-\mathop{\sum_{(i,j,k)=(s,t,u),}}_{(t,u,s),(u,s,t)}{r^{(1)}_8(x_i,x_j,x_k)\frac{f_6^{(1,2)}(x_i,x_j,x_k)}{\sqrt{x_ix_j(x_ix_j+4x_k)}}}\right\},\label{eq:oneloopFFs}
\end{eqnarray}
where the basis of rational coefficients is as follows:
\begin{eqnarray}\label{eq:ratcoeff}
 r^{(1)}_1(x_i,x_j,x_k)&=&2(x_i-4)\left(\frac{x_i}{x_k}-\frac{x_k}{x_j}\right)
\nonumber\\
r^{(1)}_2(x_i,x_j,x_k)&=&-2\left(\frac{4-x_i}{x_i} + \frac{x_i (x_j^3 - 6 x_j x_k + x_k^3)}{x_j^2 x_k^2}\right),\nonumber\\
r^{(1)}_3(x_i,x_j,x_k)&=&\frac{32x_k^2-x_ix_j(x_i^2+x_j^2-16x_k)}{x_k^2},\nonumber\\
r^{(1)}_4(x_i,x_j,x_k)&=&\frac{2x_i(x_i-4)(x_j-x_k)}{x_jx_k},\nonumber\\
r^{(1)}_5(x_i,x_j,x_k)&=&\frac{2(2x_jx_k(x_j-x_k)+x_i(x_j^3-x_k^3))}{x_j^2x_k^2},\nonumber\\
r^{(1)}_6(x_i,x_j,x_k)&=&\frac{2(x_i^2+x_j^2-4x_k)}{x_k^2},\nonumber\\
r^{(1)}_7(x_i,x_j,x_k)&=&\frac{x_i(4(x_j^2-x_i^2)+x_j(x_i^2+x_j^2)}{x_k^2},\nonumber\\
r^{(1)}_8(x_i,x_j,x_k)&=&\frac{2(x_ix_j+2x_k)}{x_k}.
\end{eqnarray}
The one-loop master integrals, defined in appendix \ref{sec:masterintegrals1L}, can be expressed through GPLs:
\begin{eqnarray}
f_2^{(1,1)}(x_s,x_t,x_u)&=&\left(-1\right)^{
\theta(x_s-4)
}\left[-G(-1;w)+G(-1;z)+G(1;w)-G(1;z)\right],\nonumber\\
f_4^{(1,2)}(x_s,x_t,x_u)&=&\left(G(1;w)-G(-1;w)\right)\left(G(-1;z)-G(1;z)\right)\nonumber\\
&&+\sum_{i,j=0}^{1}{(-1)^{i+j}\left[G((-1)^i,(-1)^j;w)+G((-1)^i,(-1)^j;z)\right]},\nonumber\\
f_6^{(1,2)}(x_s,x_t,x_u)&=&\left(-1\right)^{
\theta(x_s)+\theta(x_t)
}\left[G(0;w)\left[G(1;z)-G(-1;z)\right]+G(0;z)\left[G(-1;w)-G(1;w)\right]\right.\nonumber\\
&&+~G(1,0;w)-G(-1,0;w)+G(0,-1;w)-G(0,1;w)\nonumber\\
&&\left.-~G(1,0;z)+G(-1,0;z)-G(0,-1;z)+G(0,1;z)\right].\label{eq:oneloopfinGs}
\end{eqnarray}
The prefactors $\left(-1\right)^{
\theta(x_s-4)
}$ and $\left(-1\right)^{
\theta(x_s)+\theta(x_t)
}$ ensure the correct analytic continuation when using the convention in section \ref{sec:wzdef}, so that
the multi-valued square-root function in eq.\eqref{eq:oneloopFFs} takes the values in the first Riemann sheet.

Alternatively, the one-loop helicity amplitudes with a given massive fermion $f$ loop $\mathcal{M}_{\lambda_1\lambda_2\lambda_3\lambda_4}^{(0,0,f)}$ are
\begin{eqnarray}
-i\mathcal{M}^{(0,0,f)}_{++++}&=&8N_{c,f}Q_f^4\alpha^2\left[1-4\mathop{\sum_{(i,j,k)=(s,t,u),}}_{(t,u,s),(u,s,t)}{\frac{f_6^{(1,2)}(x_i,x_j,x_k)}{\sqrt{x_ix_j(x_ix_j+4x_k)}}}\right],\nonumber\\
-i\mathcal{M}^{(0,0,f)}_{-+++}&=&8N_{c,f}Q_f^4\alpha^2\left[1+2\left(x_s^{-1}+x_t^{-1}+x_u^{-1}\right)\mathop{\sum_{(i,j,k)=(s,t,u),}}_{(t,u,s),(u,s,t)}{f_4^{(1,2)}(x_i,x_j,x_k)}\right.\nonumber\\
&&\qquad \qquad \qquad \left.-\mathop{\sum_{(i,j,k)=(s,t,u),}}_{(t,u,s),(u,s,t)}{r^{(1)}_8(x_i,x_j,x_k)\frac{f_6^{(1,2)}(x_i,x_j,x_k)}{\sqrt{x_ix_j(x_ix_j+4x_k)}}}\right],\nonumber\\
-i\mathcal{M}^{(0,0,f)}_{--++}&=&8N_{c,f}Q_f^4\alpha^2\left[-1+r_9^{(1)}(x_t,x_u,x_s)\frac{f_2^{(1,1)}(x_t,x_u,x_s)}{\sqrt{x_t(x_t-4)}}+r_9^{(1)}(x_u,x_t,x_s)\frac{f_2^{(1,1)}(x_u,x_t,x_s)}{\sqrt{x_u(x_u-4)}}\right.\nonumber\\
&&-r_{10}^{(1)}(x_s,x_t,x_u)\left(f_4^{(1,2)}(x_t,x_u,x_s)+f_4^{(1,2)}(x_u,x_s,x_t)\right)+2(x_s-2)\frac{f_6^{(1,2)}(x_s,x_t,x_u)}{\sqrt{x_sx_t(x_sx_t+4x_u)}}\nonumber\\
&&\left.+2(x_s-2)\frac{f_6^{(1,2)}(x_u,x_s,x_t)}{\sqrt{x_sx_u(x_sx_u+4x_t)}}-r_{11}^{(1)}(x_s,x_t,x_u)\frac{f_6^{(1,2)}(x_t,x_u,x_s)}{\sqrt{x_tx_u(x_tx_u+4x_s)}}\right],\label{eq:f1Lamp}
\end{eqnarray}
where
\begin{eqnarray}
r_9^{(1)}(x_i,x_j,x_k)&=&\frac{(x_i-4)(x_i-x_j)}{x_k},\nonumber\\
r_{10}^{(1)}(x_i,x_j,x_k)&=&1-\frac{4}{x_i}-\frac{2x_jx_k}{x_i^2},\nonumber\\
r_{11}^{(1)}(x_i,x_j,x_k)&=&4-2x_i-x_jx_k+\frac{2x_jx_k(x_jx_k+4x_i)}{x_i^2}.
\end{eqnarray}
It is interesting to note that, while the conversion between the helicity amplitudes and the form factors is merely a linear transformation (see eq.\eqref{eq:ampinff}), the expressions for the amplitudes are shorter than those for the form factors. Such an observation motivates us to simplify at the level of the helicity amplitude rather than at the form factor. Note that the one-loop amplitudes have been given in ref.~\cite{Bardin:2009gq} in terms of one-loop scalar integrals, rather than as presented here in terms of a UT basis. The UT basis has the advantage of rendering shorter expressions. Our one-loop results have been numerically checked against \madloop~\cite{Hirschi:2011pa,Alwall:2014hca}.

\subsection{Two-loop helicity amplitudes}
\label{sec:twoloopamp}

We are now ready to present the UV-renormalised two-loop helicity amplitudes~\footnote{At this order, we only need to renormalise the internal fermion mass $m_f$. In this article, we choose the on-shell renormalisation condition for the mass.} with the full fermion mass dependence. Following the IBP reduction, we obtain large rational coefficients in front of the master integrals. We put dedicated efforts to simplify these coefficients into more compact and well-organised forms as given below. This is achieved with the help of \finiteflow~\cite{Peraro:2019svx} and \multivariateapart~\cite{Heller:2021qkz}. Also, the relations between the master integrals at different weights as given in eqs.\eqref{eq:fnrelation1},\eqref{eq:fnrelation2} and \eqref{eq:fnrelation3} were one of the crucial ingredients for simplifying the final expressions. The size of the amplitude file is reduced from $\sim 300$ MB just after IBP to around $180$ KB counting with {\tt Mathematica}. 

After simplification, the two-loop QCD helicity amplitudes are given as follows, with $C_{F,f}=(N_{c,f}^2-1)/(2N_{c,f})$. The QED counterparts are obtained with a simple rescaling  $\mathcal{M}^{(0,1,f)}_{\lambda_1\lambda_2\lambda_3\lambda_4}/\mathcal{M}^{(1,0,f)}_{\lambda_1\lambda_2\lambda_3\lambda_4}=\alpha Q_f^2/(\alpha_s C_{F,f})$. It is implicitly understood that the two-loop QCD amplitudes only apply to quarks, while the QED amplitudes are non-zero for all charged fermions, \ie, quarks and charged leptons in the SM. 

The all-plus and one-minus amplitudes are
\begin{align}
i\mathcal{M }^{(1,0,f)}_{++++}& = 4 N_{c,f} Q_f^4 \alpha^2 \frac{\alpha_s}{\pi} C_{F,f}   \left\{3 -4
 \mathop{\sum_{(i,j,k)=(s,t,u),}}_{(t,u,s),(u,s,t)} \left[
 \left(\frac{2}{x_i} + {2\over x_k} - {1\over x_j}\right) f_5^{(1,2)}( x_i,x_j, x_k)
 \right. \right.
 \nonumber \\ 
- &{1\over x_i^2} f_{24}^{(2,4)}(x_i,x_j, x_k) 
 +\frac{2}{ \sqrt{x_i x_j(x_i x_j+4 x_k)}}\left(\frac{x_i x_j+4 x_k}{x_k}f_{19}^{(2,2)}(x_i, x_j,x_k)  + 
    f_{22}^{(2,3)}(x_i, x_j, x_k)\right) 
 \nonumber \\ 
+ &\frac{1 }{2x_ix_j} 
  \left(6f_3^{(2,3)}( x_i, x_j, x_k) - 
      f_{14}^{(2,3)}(x_k, x_i, x_j) - 
     {4} f_{16}^{(2,3)}(x_k, x_i, x_j)
    -   {8} f_{20}^{(2,3)}(x_i, x_j, x_k) \right.
     \nonumber \\ & \left. \left.
    +2 f_{15}^{(2,2)}(x_k, x_i, x_j) f_3^{(1,1)}(x_k, x_i, x_j)
    \right)     \right]
    \nonumber \\ &
 + \left.
 \mathop{\sum_{(i,j,k)=(s,t,u),(s,u,t),(t,s,u),}}_{(t,u,s),(u,s,t),(u,t,s)}
 \left[
  \frac{8 ~ f_{19}^{(2,3)}(x_i, x_j, x_k) }{\sqrt{x_i x_j(x_i x_j+4 x_k)}} 
    -\frac{2 (x_i-2)  f_{27}^{(2,4)}( x_i, x_j, x_k)  }{\sqrt{x_i(x_i-4)}\sqrt{x_i x_j(x_i x_j+4 x_k)}} 
    \right] \right\}\,,\label{eq:Mpppp2L}
\end{align}
and
\begin{align}
 i\mathcal{M}^{(1,0,f)}_{-+++} &=  N_{c,f} Q_f^4 \alpha^2 \frac{\alpha_s}{\pi} C_{F,f}   \left\{ \mathop{\sum_{(i,j,k)=(s,t,u),}}_{(t,u,s),(u,s,t)} \left[\frac{1}{\sqrt{x_j (x_j-4)}} \left(  2r_1^{(1)}(x_j,x_k,x_i)
        f_3^{(1,1)}(x_i,x_j,x_k)
        \right.\right.\right.
        \nonumber \\ &
        \left.+
 r_{1}^{(2)}(x_i,x_j,x_k) \Big(f_3^{(1,2)}(x_i,x_j, x_k) -\frac{ 1}{6} f_{15}^{(2,2)}(x_i,x_j,x_k) \Big) \right)
 -\frac{2 x_i x_j f_6^{(1,2)}( x_i,x_j,x_k)}{\sqrt{x_i x_j (x_i x_j+4 x_k)}} 
\nonumber \\ &
+  r_2^{(2)}( x_i,x_j,x_k) \left(f_{3}^{(2,3)}( x_i,x_j,x_k)+{1\over 3} f_3^{(1,1)}(x_k,x_i,x_j) f_{15}^{(2,2)}(x_k,x_i,x_j)\right)
\nonumber \\ &
+ 
 \frac{ r_{3}^{(2)}(x_i,x_j,x_k)}{\sqrt{x_i (x_i-4)}} f_5^{(2,3)}(x_i,x_j,x_k)
 + 
 \frac{ r_{4}^{(2)}(x_i,x_j,x_k)}{\sqrt{x_i (x_i-4)}} f_{13}^{(2,4)}( x_i,x_j,x_k)\nonumber\\& 
  + r_{5}^{(2)}(x_i,x_j,x_k) f_{14}^{(2,3)}( x_i,x_j,x_k )
   +r_6^{(2)}( x_i, x_j, x_k )
      f_5^{(1,2)}(x_i,x_j,x_k) \nonumber\\&
    +r_{7}^{(2)}(x_i,x_j,x_k)
   f_5^{(1,3)}( x_i, x_j, x_k )
      -\left({6\over x_i} +{4\over x_j} + {6\over x_k}\right) f_{17}^{(2,4)}(x_i, x_j, x_k)
       \nonumber \\ &
   + \left({ 4\over x_i} + {4\over x_j} + {8\over x_k}\right)\left(
     f_{21}^{(2,4)}(x_i,x_j,x_k)-
     {1\over 2} f_{24}^{(2,4)}(x_k,x_i,x_j) \right)
    +
  r_8^{(2)}(x_i,x_j,x_k) f_{20}^{(2,3)}(x_i,x_j,x_k)      
 \nonumber \\ &
 \left.
+
 r_{10}^{(2)}(x_i,x_j,x_k)\frac{f_{19}^{(2,2)}(x_i,x_j,x_k) }{\sqrt{x_i x_j (x_i x_j+4 x_k)}}
-
\left(32 + \frac{16 x_i x_j}{x_k}\right) \frac{
    f_{22}^{(2,3)}(x_i,x_j,x_k)}{\sqrt{x_i x_j (x_i x_j+4 x_k)}}
    \right]  
   \nonumber \\ &
  - \mathop{\sum_{(i,j,k)=(s,t,u),(s,u,t)}}_{(t,u,s),(t,s,u),(u,s,t),(u,t,s)}\left[
     \left({4\over x_i} +{6\over x_j} + {6\over x_k}\right)
     f_{26}^{(2,4)}(x_i,x_j,x_k)
      + \left({2\over x_j} -{2\over x_k} \right)
     f_{29}^{(2,4)}(x_i, x_j,x_k) \right.
  \nonumber \\ &
    - 
\frac{r_{9}^{(2)}(x_i,x_j,x_k) f_{18}^{(2,3)}(x_i,x_j,x_k) }{\sqrt{x_i \left(x_i (x_j-1 )^2 - 4 x_j^2\right)}} 
- \left(32 + \frac{16 x_i x_j}{x_k}\right)\frac{
     f_{19}^{(2,3)}(x_i,x_j,x_k) }{\sqrt{x_i x_j (x_i x_j+4 x_k)}}
    \nonumber \\ &
   \left. -
  r_{11}^{(2)}(x_i,x_j,x_k)  \frac{f_{25}^{(2,4)}(x_i,x_j,x_k) }{\sqrt{x_i x_j (x_i x_j+4 x_k)}}
 +\left(2+ \frac{ x_i x_j}{x_k}\right)\frac{4 (x_i-2)}{\sqrt{x_i (x_i-4)}} \frac{f_{27}^{(2,4)}(x_i,x_j,x_k)}{\sqrt{x_i x_j (x_i x_j+4 x_k)}}\right.
   \nonumber \\ & 
\left.\left. -\left( \frac{4(x_i-2)}{x_k}-\frac{2x_k}{x_j}\right) \frac{f_{28}^{(2,4)}(x_i,x_j,x_k)}{\sqrt{x_i (x_i-4)} }
     \right]\right\}\,,\label{eq:Mmppp2L}
\end{align}
where the rational coefficients are defined as
\begin{align}
    r_{1}^{(2)}(x_i,x_j,x_k) &= -3(x_j-4) \left(
    \frac{ x_i-x_k-x_i x_j}{x_j (x_i-1)^2-4 x_i^2}
    +\frac{x_k-x_i-x_k x_j}{ x_j (x_k-1)^2 -4 x_k^2}
    \right)\,,
    \nonumber \\  
r_2^{(2)}( x_i,x_j,x_k) &=
    {2\over x_i}+{9\over x_j}+ {30\over x_k}+\frac{36}{x_i x_j}
    -\frac{8}{x_i x_k}
    +\frac{20+x_i-2x_j}{x_ix_k+4 x_j}+\frac{112+15x_i+13x_j}{2(x_i x_j+4 x_k)}\,,
     \nonumber \\ 
 r_3^{(2)}(x_i,x_j,x_k)&=-\frac{
2 (x_i-4)^2 (x_k-x_j)^2}{(x_i x_j +4 x_k)(x_ix_k+4 x_j) }\,,
\nonumber \\  
 r_{4}^{(2)}(x_i,x_j,x_k)&=
2 (x_i-4) \frac{x_j^2+x_k^2}{x_i x_jx_k}\,,
\nonumber \\  
 r_{5}^{(2)}(x_i,x_j,x_k) &=-{1\over 12}\left({93\over x_i} + {68\over x_j} +{135\over x_k}+
\frac{88}{x_i x_j} + \frac{168}{x_i x_k} + \frac{
 116+8x_i+21 x_j}{ (x_i x_j + 4 x_k)} + \frac{44+x_i+13x_j}{
 x_j x_k+4 x_i }\right)\,,
\nonumber \\  
r_{6}^{(2)}(x_i,x_j,x_k) &= -4+{4\over x_j}+\frac{4
 x_j (x_i^3 - 2 x_ix_k + x_k^3)}{
 x_i^2 x_k^2}
 + \frac{2+x_i+x_k(x_j-3)}{x_j (x_k-1)^2-4 x_k^2}   
 + \frac{2+x_k+x_i(x_j-3)}{
 x_j  (x_i-1)^2-4 x_i^2}
 \nonumber \\ &
 +
\frac{16-6 x_i-6 x_j (x_k-3) }{ x_k (x_j-1)^2- 4 x_j^2} + \frac{16-6 x_k-6 x_j (x_i-3)}{x_i (x_j-1)^2-4 x_j^2}\,,
\nonumber \\
r_7^{(2)}( x_i, x_j, x_k ) &= -6 r_5^{(2)}( x_i, x_j, x_k )-32 \left(x_i^{-1}+x_j^{-1}+x_k^{-1}\right)\,,
\nonumber \\  
r_8^{(2)}(x_i,x_j,x_k) & ={8\over x_i}+{8\over x_j}-{34\over x_k}-{88\over x_i x_j}
 - \frac{72 - 9 x_k}{x_i x_j + 4 x_k}\,,
\nonumber \\  
r_{9}^{(2)}(x_i,x_j,x_k) & =
37-8x_i+\frac{(17-9 x_i)x_i-8}{x_k}-\frac{2 (4+x_k)}{x_j}-\frac{x_k (28+9 x_i)-4 x_i}{x_i x_j+4 x_k}
-\frac{2 x_i(1+x_j)}{x_i (x_j-1)^2-4x_j^2}\,,
\nonumber \\  
r_{10}^{(2)}(x_i,x_j,x_k) & =
 56+\frac{4x_ix_j (x_ix_j-(x_k-4) x_k)}{x_k^2}+\frac{4x_j (1+2 x_i)}{ x_j(x_i-1)^2-4 x_i^2}+\frac{4x_i(1+2x_j)}{x_i(x_j-1)^2-4 x_j^2}\,,
 \nonumber \\ 
r_{11}^{(2)}(x_i,x_j,x_k) &=
 4 (1-x_j)-\frac{2 x_i^2}{x_k}-\frac{4 (x_ix_k+4 x_k)}{x_i x_j+4 x_k}.
\end{align}
They are the shortest and second-shortest amplitudes, and are written in a form that manifests the $s,t,u$ symmetry. It is also interesting to note that in the high-energy limit, \ie, $x_s,-x_t,-x_u\gg 1$, the all-plus amplitude should approach to a constant as shown in ref.~\cite{Bern:2001dg}. We can indeed verify that eq.\eqref{eq:Mpppp2L} is only left with the first constant term that matches with ref.~\cite{Bern:2001dg}.

The most-lengthy amplitudes are the two-minus-two-plus helicity amplitudes. Thanks to the relation in eq.\eqref{eq:Mmmpprelation}, we only need to present $\mathcal{M}^{(1,0,f)}_{--++}$, which is:
\begin{align}
 i\mathcal{M}^{(1,0,f)}_{--++} &=  N_{c,f} Q_f^4 \alpha^2 \frac{\alpha_s}{\pi} C_{F,f}   \left\{4 
+\left({8\over x_s} + {4x_s\over x_tx_u}\right)\left[\mathop{\sum_{(i,j,k)=(s,t,u),}}_{(t,u,s),(u,s,t)}\frac{x_j(x_j-4)f_3^{(1,1)}(x_i,x_j,x_k)}{\sqrt{x_j (x_j-4)}}\right]\right.
\nonumber \\ &
 +r_{17}^{(2)}(x_u,x_s,x_t)f_{5}^{(1,2)}(x_u,x_s,x_t)
-\frac{8 x_s^2 (x_s^2-16)}{(x_s x_t + 4 x_u) (x_sx_u+4 x_t)}
\frac{f_5^{(2,3)}(x_s,x_t,x_u)}{\sqrt{x_s(x_s-4)}}
 \nonumber \\ &
+r_{29}^{(2)}(x_s,x_t,x_u)\left(
 f_3^{(2,3)}(x_s,x_t,x_u) +{1\over 3} f_{15}^{(2,2)}(x_u,x_s,x_t) 
 f_{3}^{(1,1)}(x_u,x_s,x_t) \right)
 \nonumber \\ &
 -8 (2x_s+x_tx_u)
\frac{f_6^{(1,2)}(x_t,x_u,x_s)}{\sqrt{x_tx_u(x_tx_u+4x_s)}} 
+r_{24}^{(2)}(x_t,x_u,x_s)
\frac{f_{19}^{(2,2)}(x_t,x_u,x_s)}{\sqrt{x_t x_u (x_t x_u+4 x_s)}} 
\nonumber \\ &
+\frac{8}{x_s^2} (8+x_s-x_t^2-x_u^2) f_{21}^{(2,4)}(x_t,x_u,x_s)
-r_{25}^{(2)}(x_t,x_u,x_s)\frac{f_{22}^{(2,3)}(x_t,x_u,x_s)}{\sqrt{x_t x_u (x_t x_u+4 x_s)}}
 \nonumber \\ &
  +{16\over x_s^2}f_{24}^{(2,4)}(x_s,x_t,x_u)
+r_{26}^{(2)}(x_u,x_s,x_t)\left(
f_{14}^{(2,3)}(x_u,x_s,x_t) -6  f_5^{(1,3)}(x_u,x_s,x_t)\right)
\nonumber \\ &
+
\mathop{\sum_{(i,j,k)=(s,t,u),}}_{(s,u,t)}
\left[-16 x_i
\frac{f_6^{(1,2)}(x_i,x_j,x_k)}{\sqrt{x_ix_j(x_ix_j+4x_k)}} 
-16 (x_i-2)\frac{\left(f_{19}^{(2,3)}(x_i,x_j,x_k)-f_{22}^{(2,3)}(x_i,x_j,x_k)\right)}{\sqrt{x_i x_j (x_i x_j+4 x_k)}}\right.
\nonumber \\ &
+\frac{4 (x_i-2)^2}{\sqrt{x_i (x_i-4)}}\frac{f_{27}^{(2,4)}(x_i,x_j,x_k)}{\sqrt{x_i x_j (x_i x_j+4 x_k)}}
+r_{12}^{(2)}(x_i,x_j,x_k) \left(f_{17}^{(2,4)}( x_i, x_j,x_k)+ f_{26}^{(2,4)}(x_k,x_j,x_i)\right) 
\nonumber \\ &
+r_{13}^{(2)}(x_i,x_j,x_k) f_{20}^{(2,3)}(x_i,x_j,x_k)
+r_{16}^{(2)}(x_i,x_j,x_k)f_{5}^{(1,2)}(x_i,x_j,x_k) \nonumber\\&
+r_{18}^{(2)}(x_i,x_j,x_k) \frac{f_{3}^{(1,2)}(x_i,x_j,x_k) }{\sqrt{x_j(x_j-4)}}
+r_{20}^{(2)}(x_i,x_j,x_k)
\frac{f_{15}^{(2,2)}(x_i,x_j,x_k)}{\sqrt{x_j(x_j-4)}} \nonumber\\&
+\frac{8 x_i (2 x_i+(x_i-4) x_j^2)}{x_j (x_i x_j+4 x_k)}\frac{f_{18}^{(2,3)}(x_i,x_j,x_k)}{\sqrt{x_i (x_i (x_j-1)^2 - 4 x_j^2)}}
\nonumber \\ &
+r_{23}^{(2)}(x_i,x_j,x_k)
\frac{f_{19}^{(2,2)}(x_i,x_j,x_k)}{\sqrt{x_i x_j (x_i x_j+4 x_k)}} 
+\frac{16 x_i^2}{x_i x_j + 4 x_k}\frac{f_{25}^{(2,4)}(x_i,x_j,x_k)}{\sqrt{x_i x_j (x_i x_j+4 x_k)}}
\nonumber \\ &
+r_{27}^{(2)}(x_i, x_j, x_k)f_{14}^{(2,3)}(x_i,x_j,x_k)
+r_{28}^{(2)}(x_i,x_j,x_k) f_5^{(1,3)}(x_i,x_j,x_k)
\nonumber \\ &\left.
+r_{31}^{(2)}(x_i,x_j,x_k)
 f_{15}^{(2,2)}(x_i,x_j,x_k)f_{3}^{(1,1)}(x_i,x_j,x_k)
\right]
\nonumber \\ & 
+
\mathop{\sum_{(i,j,k)=(u,s,t),}}_{(t,s,u)}\left[
-16 (x_j-2)\frac{f_{19}^{(2,3)}(x_i,x_j,x_k)}{\sqrt{x_i x_j (x_i x_j+4 x_k)}}
+r_{21}^{(2)}(x_i,x_j,x_k)
\frac{f_{18}^{(2,3)}(x_i,x_j,x_k)}{\sqrt{x_i (x_i (x_j-1)^2 - 4 x_j^2)}} \right.
\nonumber \\ & 
-\frac{8}{x_j^2} (4 + x_j + 3 x_k - x_i^2)\left(f_{26}^{(2,4)}(x_i,x_j,x_k) 
+\frac{1}{2} f_{29}^{(2,4)}(x_i,x_j,x_k)\right)
\nonumber \\ &
+\frac{4 (x_j-2) (x_i-2)}{\sqrt{x_i (x_i-4)}}\frac{f_{27}^{(2,4)}(x_i,x_j,x_k)}{\sqrt{x_i x_j (x_i x_j+4 x_k)}} 
+4 (4 x_j+12 x_i-x_i^3-x_k^2)\frac{f_{28}^{(2,4)}(x_i,x_j,x_k)}{x_j^2\sqrt{ x_i ( x_i-4)}}
\nonumber \\ &\left.
+r_{30}^{(2)}(x_i,x_j,x_k)f_{3}^{(2,3)}(x_i,x_j,x_k)
\right]
\nonumber \\ &
+
\mathop{\sum_{(i,j,k)=(t,u,s),}}_{(u,t,s)}\left[
 r_{14}^{(2)}(x_i,x_j,x_k)f_{20}^{(2,3)}(x_i,x_j,x_k)
 +\frac{r_{25}^{(2)}(x_i,x_j,x_k)}{\sqrt{x_i x_j (x_i x_j+4 x_k)}}\left(f_{19}^{(2,3)}(x_i,x_j,x_k)\right.\right.
\nonumber \\ &\left.
-\frac{(x_i-2)f_{27}^{(2,4)}(x_i,x_j,x_k)}{4\sqrt{x_i(x_i-4)}}
\right)
-2(x_i^2+x_j^2-4 x_k)
\frac{(x_i-2)f_{28}^{(2,4)}(x_i,x_j,x_k)}{x_k^2\sqrt{ x_i ( x_i-4)}}
\nonumber \\ &
+r_{15}^{(2)}(x_i,x_j,x_k)\left(f_{24}^{(2,4)}(x_i,x_j,x_k)-f_{29}^{(2,4)}(x_i,x_j,x_k) \right)
+r_{19}^{(2)}(x_i,x_j,x_k)
\frac{f_{13}^{(2,4)}(x_i,x_j,x_k)}{\sqrt{x_i(x_i-4)}} 
\nonumber \\ &\left.\left.
+r_{22}^{(2)}(x_i,x_j,x_k)
\frac{f_{18}^{(2,3)}(x_i,x_j,x_k)}{\sqrt{x_i (x_i (x_j-1)^2 - 4 x_j^2)}} 
\right]
 \right\}\,.\label{eq:Mmmpp2L}
\end{align}
Here the additional rational coefficients are:
\begin{align}
    r_{12}^{(2)}(x_i,x_j,x_k) &=4 + \frac{8 (1 + x_j)}{x_i }+ \frac{8 (x_j^2+3x_j-4 )}{x_i^2}\,,
\nonumber \\
r_{13}^{(2)}(x_i,x_j,x_k) &=-12-\frac{108+24 x_j}{x_i^2}+{72+24x_j \over x_i}+{48\over x_j}
-{16\over x_k}-\frac{128}{x_i x_j}
+\frac{144-36 x_i}{x_i x_j+4 x_k}\,,
\nonumber \\
r_{14}^{(2)}(x_i,x_j,x_k) &=
24-{102\over x_k}+\frac{34 x_k-64}{x_i x_j}+\frac{36-48 x_i x_j}{x_k^2}\,,
\nonumber \\
r_{15}^{(2)}(x_i,x_j,x_k) &=
2 + \frac{4+4x_j}{x_k}+\frac{4 (x_j^2+3x_j-4)}{x_k^2}\,,
\nonumber \\
r^{(2)}_{16}(x_i,x_j,x_k)&=
 38-{48\over x_j}-\frac{152-48 x_j}{x_i}-\frac{48 +8x_j}{x_k}+{4 x_j^2\over x_k^2}
 -{8 x_j^2\over x_i^2}-\frac{4-12 x_i}{x_j(x_i-1)^2-4x_i^2}
 \nonumber \\   &
 -\frac{ 4+2x_i-2(x_j+1)x_k}{ x_j(x_k-1)^2-4x_k^2}
 -\frac{8 \left(4+5x_j-5x_i-2x_jx_k\right)}{x_k(x_j-1)^2 -4x_j^2}\,,
 \nonumber \\   
 r^{(2)}_{17}(x_i,x_j,x_k) &=
 4 \left(4+{8\over x_j}+{x_j^2\over x_k^2}+{x_j^2\over x_i^2}
 -\frac{4 (x_j-3)x_j}{x_k x_i}-\frac{4 (x_j+2)}{x_k(x_j-1)^2-4 x_j^2}
 -\frac{4 (x_j+2)}{ x_i(x_j-1)^2-4 x_j^2}\right)\,,
 \nonumber \\ 
r^{(2)}_{18}(x_i,x_j,x_k) &=x_j(x_j-4)\left(
 {40 \over x_i}+{20\over x_j}-\frac{3 (1+x_k)^2-3 x_i (3+x_k^2)}{x_i (x_j(x_k-1)^2-4 x_k^2)}
 +\frac{ 3 (x_i-1)^2(x_i+1)}{x_i (x_j(x_i-1)^2-4 x_i^2)}\right)\,,
\nonumber \\ 
r_{19}^{(2)}(x_i,x_j,x_k)&=(x_k-4)\left(2-\frac{12x_k}{x_j^2}+\frac{4(x_k-2)}{x_j}\right)\,,
\nonumber \\ 
r_{20}^{(2)}(x_i,x_j,x_k)&=
-\frac{1}{6} r_{18}^{(2)}(x_i,x_j,x_k)-\frac{8}{3x_i} (x_j-x_k)(x_j-4)\,,
\nonumber \\ 
r_{21}^{(2)}(x_i,x_j,x_k)&=
4 \left(33 - 4 x_j + 3 x_k+ x_i x_j - 2 x_i^2 - \frac{x_i (2 x_i+9)}{x_j^2} + \frac{
   x_i (4 x_i+19)}{x_j}  + \frac{2 x_i-2 x_j (x_j-x_k)}{x_i (x_j-1)^2-4 x_j^2)}\right)\,,
   \nonumber \\ 
r_{22}^{(2)}(x_i,x_j,x_k)&=
   8-2 x_k-6 x_j+2 x_k x_j+6 x_j^2+\frac{4 x_j (x_j+1)^2 (9+4 x_j)}{x_k^2}+\frac{4 (x_k-x_j) (x_k-x_j+x_j x_i)}{x_i (x_j-1)^2-4 x_j^2}
    \nonumber \\ &
   +\frac{4 (9+x_j (5 x_j^2+5 x_j-11))}{x_k}\,,
    \nonumber \\ 
   r_{23}^{(2)}(x_i,x_j,x_k)&=
   -264 - 4 x_i (x_j-2) + 8 x_j (x_j+3) + \frac{
 8 (8 x_i^2 - 3 x_j (1-x_i))}{x_j(x_i-1)^2 -4 x_i^2} - \frac{ 4 x_j^4}{x_k^2} + \frac{4 x_j^2 (x_j+14)}{x_k}\,,
 \nonumber \\ 
   r_{24}^{(2)}(x_i,x_j,x_k)&=
 4x_i x_j (4 x_k + x_i x_j)\left(
 {2\over x_k^2} + \frac{3 (x_k-6)}{x_i x_j x_k} - \frac{
    x_k-2+x_i (x_j-1)}{x_j (x_j(x_i-1)^2-4 x_i^2)}- \frac{
   x_k-2+x_j (x_i-1)}{x_i (x_i(x_j-1)^2-4 x_j^2)}\right)\,,
\nonumber \\ 
   r_{25}^{(2)}(x_i,x_j,x_k)&=
   32 - 16 x_k - 8 x_i x_j + \frac{64 x_i x_j}{x_k} + \frac{16 x_i^2 x_j^2}{x_k^2}\,,
\nonumber \\ 
r_{26}^{(2)}(x_i,x_j,x_k)&=\frac{1}{3}\left(28-{90\over x_j^2}-{74 \over x_j}+{19x_j-32 \over x_i x_k}+{12 (4-x_j) \over x_ix_j+4x_k}+{12 (4-x_j) \over x_jx_k+4x_i}\right)\,,
\nonumber \\ 
r_{27}^{(2)}(x_i,x_j,x_k)&=\frac{1}{6}\left(37+{45 \over x_j}-{32 \over x_i x_j}+{73(2x_j-1) \over x_i}+{2(40x_j^2-175x_j-63) \over x_i^2}+{x_i-32\over x_i x_k}-{16(x_i-4)\over x_ix_j+4x_k}\right)\,,
\nonumber \\ 
r_{28}^{(2)}(x_i,x_j,x_k)&=-45+{105-162x_j \over x_i}+{3 \over x_j}+{32 \over x_ix_j}+{126+350x_j-48x_j^2 \over x_i^2}-{1\over x_k}+{32\over x_ix_k}+{16(x_i-4) \over x_ix_j+4x_k}\,,
\nonumber \\ 
   r_{29}^{(2)}(x_i,x_j,x_k)&=
  8+{252 \over x_i^2}-{52 \over x_i}+{8(x_i-8) \over x_jx_k}+{30(x_i-4) \over x_i x_j+4x_k}+{30(x_i-4) \over x_i x_k+4x_j}\,,
\nonumber \\
r_{30}^{(2)}(x_i,x_j,x_k)&=-10+{29 \over x_i}-{32 \over x_ix_j}+{23-20x_k \over x_j}+{25 \over x_k}-{32 \over x_jx_k}+{8(x_j-4)\over x_ix_j+4x_k}+{90-52x_k-40x_k^2\over x_j^2}\,,
\nonumber \\
r_{31}^{(2)}(x_i,x_j,x_k)&=\frac{1}{3}\left(-26+{59-52x_j \over x_i}+{21 \over x_j}-{32\over x_ix_j}+{90+52x_j-40x_j^2 \over x_i^2}+{25\over x_k}-{32 \over x_i x_k}+{8(x_i-4)\over x_ix_j+4x_k}\right)\,.
\end{align}
The amplitude $\mathcal{M}^{(1,0,f)}_{--++}$ is symmetric in $x_t$ and $x_u$. The expressions of all these two-loop amplitudes are also provided in the ancillary files. We have also implemented them into a {\tt Fortran} code, that will be published elsewhere. The timing of evaluating the 2-loop amplitude square after summing all helicity configurations is around $0.3$ second for a random phase space point on a single core with the 2.6 GHz Intel Core i7 CPU.

We have also carried out a few checks. First, the crossing symmetry relations of the amplitudes have been explicitly verified by plugging in the concrete analytic expressions. Our high-energy limit approaches the massless results in ref.~\cite{Bern:2001dg} numerically, and our low-energy limit approaches the results present in ref.~\cite{Martin:2003gb} (also see appendix \ref{sec:ampLElimit}). Finally, at the cross section level, our results agree perfectly with the pure numerical approach based on the local unitarity construction~\cite{Capatti:2020xjc,Capatti:2022tit}.


\section{Conclusion}
\label{sec:conclusion}

In this paper, we have presented the compact and analytic two-loop helicity amplitudes of the LbL process $\gamma\gamma\to\gamma\gamma$ keeping the full dependence of the internal fermion mass. The results remain simple enough even at two loops to be written in a few pages thanks to the numerous symmetries of the process. They could serve as a benchmark for developing novel multi-loop techniques, such as the recent local unitarity approach~\cite{Capatti:2020xjc,Capatti:2022tit}. The connection between the simplicity of our amplitudes and the string-derived Bern-Kosower formalism~\cite{Bern:1990cu,Bern:1991aq,Strassler:1992zr} needs to be explored in the future. The phenomenological applications and the comparisons with the local unitarity method will be reported in a companion paper~\cite{H:2023znv}. As an outlook, it is intriguing in the future to take into account the full electroweak corrections and next-to-next-to-leading order QCD and QED corrections. Their phenomenological relevance may show up in UPCs at the high-luminosity LHC by tagging proton(s) with forward proton spectrometers and future $e^+e^-$ colliders.

\section*{Acknowledgements}
We thank Claude Duhr, Mathijs Fraaije, Johannes Henn, Valentin Hirschi, Kirill Melnikov, Lukas Simon, Vasily Sotnikov, Guoxing Wang and Simone Zoia for useful discussions. This work is supported by the grants from the ERC (grant 101041109 `BOSON', grant 101043686 `LoCoMotive'), the French ANR (grant ANR-20-CE31-0015 `PrecisOnium'), the French LIA FCPPL, the European Union's Horizon 2020 research and innovation program (grant 824093 STRONG-2020, EU Virtual Access `NLOAccess').

\newpage
\appendix

\section{Kinematic regions and analytic continuation}
\label{sec:wzdef}

In this section, we determine the kinematic regions that will be needed in two-loop amplitude calculations. In particular, we use the following parametrisation for the Mandelstam variables for all the polylogarithmic integrals:
\begin{eqnarray}
\label{eq:parametrisation2}
\frac{s_{ij}}{m_f^2}&=&x_{ij}=-\frac{4(w-z)^2}{(1-w^2)(1-z^2)},\quad \frac{s_{jk}}{m_f^2}=x_{jk}=-\frac{(w-z)^2}{wz}, \quad s_{ik}=-s_{ij}-s_{jk}.
\end{eqnarray}
In the above parametrisation, there are multiple solutions of $w$ and $z$ in terms of $x_{ij}$ and $x_{jk}$. We choose a unique solution that covers the whole $2\to 2$ scattering phase space. First of all, since there is a symmetry in $w$ and $z$, we assume $\Re{z}\geq \Re{w}$ if $w,z\in\mathbb{R}$. we derive the regions of $w$ and $z$ that cover the required three kinematic regions:
\begin{itemize}
\item Region I: $s_{ij}<0,s_{jk}<0,s_{ik}>0$ $\longrightarrow$ $0<w\leq z<1$.
\item Region II: $s_{ij}>0,s_{jk}<0,s_{ik}<0$ $\longrightarrow$ this region can be further divided into 2 sub-regions.
\begin{itemize}
\item $s_{ij}>4m_f^2,s_{jk}<0,s_{ik}<0$: $\sqrt{2}-1<w<1<z<\frac{1}{w}$ or $0<w<\sqrt{2}-1,1<z<\frac{1+w}{1-w}$. It can also be written as $0<w<1<z<{\rm min}(\frac{1}{w},\frac{1+w}{1-w})$.
\item $0<s_{ij}<4m_f^2,s_{jk}<0,s_{ik}<0$: $w,z\in \mathbb{C}$, $z=\frac{w}{|w|^2}$. $\sqrt{2}-1<\Re{w}<1,0<\Im{w}<\sqrt{1-\left(\Re{w}\right)^2}$ or $0<\Re{w}<\sqrt{2}-1,\sqrt{1-2\Re{w}-\left(\Re{w}\right)^2}<\Im{w}<\sqrt{1-\left(\Re{w}\right)^2}$. It can also be rewritten as $0<\Re{w}<1$ and $\sqrt{{\rm max}(0,1-2\Re{w}-\left(\Re{w}\right)^2)}<\Im{w}<\sqrt{1-\left(\Re{w}\right)^2}$.
\end{itemize}
\item Region III: $s_{ij}<0,s_{jk}>0,s_{ik}<0$  $\longrightarrow$ this region can also be further divided into 2 sub-regions.
\begin{itemize}
\item $s_{ij}<0,s_{jk}>4m_f^2,s_{ik}<0$: $1-\sqrt{2}<w<0<z<-w$ or $-1<w<1-\sqrt{2},0<z<\frac{1+w}{1-w}$. It can also be written as $-1<w<0<z<{\rm min}(-w,\frac{1+w}{1-w})$.
\item $s_{ij}<0,0<s_{jk}<4m_f^2,s_{ik}<0$: $w,z\in \mathbb{C}$, $z=-w^*$. $1-\sqrt{2}<\Re{w}<0,0<\Im{w}<\sqrt{1+2\Re{w}-\left(\Re{w}\right)^2}$.
\end{itemize}
\end{itemize}
The regions of $(w,z)$ in the above cover the entire phase space only once. 

For $w,z\in\mathbb{R}$, we need to know the sign of their imaginary parts in order to choose the correct Riemann sheet in GPLs. Since  $s_{ij}\to s_{ij}+i0, s_{jk}\to s_{jk}+i0, s_{ik}\to s_{ik}+i0$ and $m_f^2\to m_f^2-i0$, it is sufficient for our purpose to take $w\to w-i0$. For the square roots in the prefactors of the canonical basis eq.\eqref{eq:UTbasis4mf}, we always evaluate them in the first Riemann sheet for simplicity. However, a minus sign stemming from the square roots should be introduced in the canonical master integrals $f_n^{(2,\omega)}(x_{ij},x_{jk},x_{ik})$ with the corresponding square roots when the following condition is satisfied
\begin{eqnarray}
\left\{
\begin{array}{ll}
\sqrt{s_{ij}(s_{ij}-4m_f^2)} &~~{\rm with}~s_{ij}>4m_f^2\\
\sqrt{s_{jk}(s_{jk}-4m_f^2)} &~~{\rm with}~0<s_{jk}<4m_f^2\\
\sqrt{s_{ij}s_{jk}(s_{ij}s_{jk}-4m_f^2(s_{ij}+s_{jk}))} &~~{\rm with}~s_{ij}>0~{\rm or}~s_{jk}>0 .
\end{array}\right. \,
\end{eqnarray}
For instance, for the two square roots $\sqrt{s_{ij}(s_{ij}-4m_f^2)}$ and $\sqrt{s_{ij}s_{jk}(s_{ij}s_{jk}-4m_f^2(s_{ij}+s_{jk}))}$ in $f_{27}^{(2,4)}$, we should multiply the prefactor $(-1)^{\theta(x_{ij}-4)}(-1)^{\theta(x_{ij})+\theta(x_{jk})}$ in front of the expression consisting of $G$'s as what is similarly done in eq.\eqref{eq:oneloopfinGs}. We have taken care of these prefactors in the relations given in eqs.\eqref{eq:fnrelation2} and \eqref{eq:fnrelation3}, and the remaining square roots in the amplitudes can be evaluated on the first Riemann sheet. The same applies to the one-loop master integrals $f_n^{(1,\omega)}(x_{ij},x_{jk},x_{ik})$, which are defined in appendix \ref{sec:masterintegrals1L}.

\section{Analytic boundary conditions\label{app:BCs}}

In this section, we present the analytic boundary conditions in the regions II and III defined in appendix~\ref{sec:wzdef}. For the region I, we simply  choose $x_s=x_t=x_u=0$ as the boundary point, where all the master integrals are zero except the first one $f_1^{(2,\omega)}$. In the regions II and III, we use the parametrisation $x_s=s/m_f^2=-(y_s-1)^2/y_s>0$ to rationalise the square roots in the boundary conditions. The solution is
\begin{eqnarray}
y_s&=&1-x_s/2-\sqrt{x_s(x_s-4)}/2.\label{eq:ystsol}
\end{eqnarray}
Given the analytic continuation of $x_s\to x_s+i0$, we have $y_s\to y_s-i0$ when $x_s>4$. When $0<x_{s}<4$, we simply take the square root in the first Riemann sheet in eq.\eqref{eq:ystsol} with the analytic continuation $x_{s}(x_{s}-4)\to x_{s}(x_{s}-4)+i0$. We write down the independent non-zero boundary constants that are needed for evaluating the five master integrals $f_{18}^{(2,3)},f_{25}^{(2,4)},f_{26}^{(2,4)},f_{28}^{(2,4)},f_{29}^{(2,4)}$ with the non-rationalised square root $\rho$. They are:
\begin{eqnarray}
f_2^{(2,1)}(x_s,0,-x_s)&=&G(0;y_s),\quad f_2^{(2,2)}(x_s,0,-x_s)=-\zeta_2+G(0,0;y_s)-2G(-1,0;y_s),\nonumber\\
f_4^{(2,3)}(x_s,0,-x_s)&=&3\zeta_3+\zeta_2G(0;y_s)-G(0,0,0;y_s)+2G(0,-1,0;y_s),\nonumber\\
f_{10}^{(2,3)}(x_s,0,-x_s)&=&4\zeta_3+2\zeta_2G(0;y_s)+G(0,0,0;y_s)-2G(0,1,0;y_s),\nonumber\\
f_{11}^{(2,2)}(x_s,0,-x_s)&=&\zeta_2+\frac{1}{2}G(0,0;y_s)-G(1,0;y_s),\nonumber\\
f_{18}^{(2.3)}(x_s,0,-x_s)&=&-3\zeta_3-\zeta_2G(0;y_s)+G(0,1,0;y_s)-G(1,0,0;y_s),\nonumber\\
f_{20}^{(2,3)}(x_s,0,-x_s)&=&\zeta_3+\zeta_2G(0;y_s)+G(0,0,0;y_s)-G(0,1,0;y_s)-G(1,0,0;y_s),\nonumber\\
f_{23}^{(2,3)}(x_s,0,-x_s)&=&-2\zeta_3+G(0,0,0;y_s)-2G(1,0,0;y_s),\nonumber\\
f_{26}^{(2,4)}(x_s,0,-x_s)&=&\frac{5}{2}\zeta_4+\zeta_2\left[G(0,0;y_s)-2G(1,0;y_s)\right]-G(0,0,1,0;y_s)-2G(0,1,0,0;y_s)\nonumber\\
&&-3G(1,0,0,0;y_s)+2G(1,0,1,0;y_s)+4G(1,1,0,0;y_s),\nonumber\\
f_{28}^{(2,4)}(x_s,0,-x_s)&=&-9\zeta_4-8\zeta_3G(0;y_s)-4\zeta_2G(0,0;y_s)+8G(0,-1,0,0;y_s)-16G(0,0,-1,0;y_s)\nonumber\\
&&-G(0,0,0,0;y_s)-4G(0,0,1,0;y_s)+14G(0,1,0,0;y_s),\nonumber\\
f_{29}^{(2,4)}(x_s,0,-x_s)&=&-5\zeta_4-2\zeta_2\left[G(0,0;y_s)-2G(1,0;y_s)\right]+2G(0,0,1,0;y_s)+4G(0,1,0,0;y_s)\nonumber\\
&&+6G(1,0,0,0;y_s)-4G(1,0,1,0;y_s)-8G(1,1,0,0;y_s),\nonumber\\
f_{14}^{(2,3)}(0,x_s,-x_s)&=&-6\zeta_3-2\zeta_2G(0;y_s)-12G(0,-1,0;y_s)+6G(0,0,0;y_s)-4G(0,1,0;y_s)\nonumber\\
&&+4G(1,0,0;y_s),\nonumber\\
f_{16}^{(2,2)}(0,x_s,-x_s)&=&-\frac{1}{2}G(0,0;y_s),\nonumber\\
f_{16}^{(2,3)}(0,x_s,-x_s)&=&-\frac{\zeta_3}{2}+\frac{\zeta_2}{2}G(0;y_s)+3G(0,-1,0;y_s)-\frac{1}{2}G(0,0,0;y_s)+G(0,1,0;y_s)\nonumber\\
&&-3G(1,0,0;y_s),\nonumber\\
f_{23}^{(2,3)}(0,x_s,-x_s)&=&-2\zeta_3+G(0,0,0;y_s)-2G(1,0,0;y_s),\nonumber\\
f_{26}^{(2,4)}(0,x_s,-x_s)&=&-6\zeta_4-4\zeta_3\left[G(0;y_s)-2G(1;y_s)\right]+2G(0,0,0,0;y_s)-4G(0,1,0,0;y_s)\nonumber\\
&&-4G(1,0,0,0;y_s)+8G(1,1,0,0;y_s).\label{eq:boundaryconstants}
\end{eqnarray}
The other non-zero integrals can be obtained by using the relations among the master integrals in eqs.\eqref{eq:fnrelation2} and \eqref{eq:fnrelation3}. Otherwise, they are either zero or not needed in the evaluations of the five master integrals with square roots. In eq.\eqref{eq:boundaryconstants}, $\zeta_n$ is the usual Riemann zeta function $\zeta(n)$. 

\section{One-loop master integrals}
\label{sec:masterintegrals1L}

Let us define the one-loop master integrals that are needed for the one-loop helicity amplitudes and the UV renormalisation counterterms. The family of the one-loop master integrals in $d=4-2\epsilon$ is
\begin{align}
\label{eq:define-int-1L}
I_{a_1,a_2,a_3,a_4}^{(1)}(s_{ij},s_{jk},s_{ik}) &= \frac{m_f^{2\epsilon}}{\Gamma(1+\epsilon)} \int \frac{\derive^d\ell_1}{i\pi^{\frac{d}{2}}} \frac{1}{D_1^{a_1} D_2^{a_2} D_3^{a_3}D_4^{a_4}},\quad a_i\in \mathbb{Z}\,.
\end{align}
The loop momentum is $\ell_1$ and $\Gamma$ is the Euler gamma function. The inverse propagators are
\begin{eqnarray}
D_1&=&\ell_1^2-m_f^2+i0^+,\quad D_2=(\ell_1+p_i)^2-m_f^2+i0^+,\nonumber\\
D_3&=&(\ell_1+p_i+p_j)^2-m_f^2+i0^+,\quad
D_4=(\ell_1+p_i+p_j+p_k)^2-m_f^2+i0^+.
\end{eqnarray}
The canonical basis with uniform transcendental weight properties is~\cite{Caron-Huot:2014lda}
\begin{eqnarray}
f_1^{(1)}&=&-2m_f^2I^{(1)}_{0,0,0,3},\nonumber\\
f_2^{(1)}&=&-\epsilon \sqrt{s_{ij}(s_{ij}-4m_f^2)}I^{(1)}_{1,0,2,0},\quad f_3^{(1)}=-\epsilon\sqrt{s_{jk}(s_{jk}-4m_f^2)}I^{(1)}_{0,1,0,2},\nonumber\\
f_4^{(1)}&=&\epsilon^2s_{ij}I^{(1)}_{1,1,1,0},\quad f_5^{(1)}=\epsilon^2s_{jk}I^{(1)}_{1,1,0,1},\nonumber\\
f_6^{(1)}&=&\frac{\epsilon^2}{2}\sqrt{s_{ij}s_{jk}\left(s_{ij}s_{jk}-4m_f^2(s_{ij}+s_{jk})\right)}I^{(1)}_{1,1,1,1}.
\end{eqnarray}
Again, let us define the vector $\vec{f}^{(1)}=(f_1^{(1)},\cdots,f_6^{(1)})$, which satisfies the $\epsilon$-form of differential equations
\begin{eqnarray}
\derive \vec{f}^{(1)}&=&\epsilon \derive A^{(1)}\vec{f}^{(1)}.
\end{eqnarray}
The matrix $A^{(1)}$ is independent of the dimensional regulator $\epsilon$. Unlike the two-loop case, in this case there are only three square roots $\sqrt{s_{ij}(s_{ij}-4m_f^2)}$, $\sqrt{s_{jk}(s_{jk}-4m_f^2)}$,
and  $\sqrt{s_{ij}s_{jk}(s_{ij}s_{jk}-4m_f^2(s_{ij}+s_{jk}))}$, that can be rationalised simultaneously with the parametrisation eq.\eqref{eq:parametrisation}. In other words, the one-loop master integrals can be expressed through GPLs up to any order in $\epsilon$.
Only $10$ $\dlog$ one-forms as already defined in eq.\eqref{eq:oneforms} appear in the matrix $A^{(1)}$:
\begin{eqnarray}
\derive A^{(1)}&=&\left(\begin{array}{cccccc} 
0 & 0 & 0 & 0 & 0 & 0\\
-\derive g_1 & \derive g_2 & 0 & 0 & 0 & 0 \\
\derive g_5 & 0 & \frac{\derive g_4}{3}+\frac{\derive g_6}{3} & 0 & 0 & 0\\
0 & -\derive g_1 & 0 & 0 & 0 & 0\\
0 & 0 & \derive g_5 & 0 & 0 & 0\\
0 & \derive g_{12} & -2\derive g_{14} & \derive g_{13} & \derive g_{13} & \frac{2\derive g_{10}}{3}+\frac{\derive g_{15}}{3}\\
\end{array}\right)\,.
\end{eqnarray}
The series expansion over the dimensional regulator $\epsilon$ in $\vec{f}^{(1)}=\sum_{\omega=0}{\epsilon^\omega \vec{f}^{(1,\omega)}}$ or equivalently $f_n^{(1)}=\sum_{\omega=0}{\epsilon^\omega f_n^{(1,\omega)}}$ enables the differential equation system
\begin{eqnarray}
\derive \vec{f}^{(1,\omega+1)}&=&\derive A^{(1)} \vec{f}^{(1,\omega)},\quad \omega\in\mathbb{N}
\end{eqnarray}
to be solved iteratively in GPLs with the boundary condition at $s_{ij}=s_{jk}=s_{ik}=0$. The analytic continuation should be performed as explained in appendix \ref{sec:wzdef}. We provide the solutions up to weight $\omega=3$ in the ancillary files.

\section{Canonical basis of two-loop master integrals}
\label{sec:canonmasterintegrals2L}
The canonical basis of the two-loop master integrals we choose here is as follows:
\begin{eqnarray}
f^{(2)}_1&=&\epsilon^2 I^{(2)}_{0,0,2,0,0,0,0,2,0},\quad
f^{(2)}_2=\epsilon^2 \sqrt{(s_{ij}-4m_f^2)s_{ij}} I^{(2)}_{ 2, 0, 1, 0, 2, 0, 0, 0, 0},\nonumber\\
f^{(2)}_3&=&-\epsilon^2 (s_{ij}-4m_f^2)s_{ij}I^{(2)}_{2, 0, 1, 0, 2, 0, 1, 0, 0},\quad f^{(2)}_4= -\epsilon^3 s_{ij} I^{(2)}_{ 2, 0, 0, 0, 1, 0, 1, 1, 
  0},\nonumber\\
f^{(2)}_5&=& \epsilon^3 s_{ij} \sqrt{(s_{ij}-4m_f^2)s_{ij}} I^{(2)}_{ 2, 0, 1, 0, 1, 0, 1, 1, 0},\quad f^{(2)}_6=\epsilon^2 s_{ij} I^{(2)}_{ 2, 0, 0, 0, 0, 0, 2, 0, 1},\nonumber\\
f_7^{(2)}&=& \epsilon^2 \sqrt{(s_{ij}-4m_f^2)s_{ij}}\left(2 I^{(2)}_{ 2, 0, 0, 0, 0, 0, 1, 0, 2} + I^{(2)}_{ 2, 0, 0, 0, 0, 0, 2, 0, 1}\right), \quad
f^{(2)}_8=\epsilon^2 m_f^2 s_{ij} I^{(2)}_{ 3, 0, 0, 0, 0, 0, 1, 1, 1},\nonumber\\
f^{(2)}_9&=& \epsilon^2 m_f^2 s_{ij} I^{(2)}_{ 1, 0, 1, 0, 0, 0, 0, 3, 1},\quad
f^{(2)}_{10}=-\epsilon^3 s_{ij} I^{(2)}_{1, 0, 1, 0, 0, 0, 0, 2, 1},\nonumber\\
f^{(2)}_{11}&=& \frac{3}{2} \epsilon^3 \sqrt{(s_{ij}-4m_f^2) s_{ij}} I^{(2)}_{ 1, 0, 1, 0, 0, 0, 0, 2,1} +  \epsilon^2 m_f^2\sqrt{(s_{ij}-4m_f^2)s_{ij}}(I^{(2)}_{1, 0, 1, 0, 0, 0, 0, 3, 1} + I^{(2)}_{ 2, 0, 1, 0, 0, 0, 0, 2, 1}),\nonumber\\
f^{(2)}_{12}&=& \epsilon^4 s_{ij} I^{(2)}_{ 1, 0, 1, 0, 0, 0,
   1, 1, 1},\quad
f^{(2)}_{13}=-\epsilon^4 s_{ij} \sqrt{(s_{ij}-4m_f^2) s_{ij}}
I^{(2)}_{ 1, 0, 1, 0, 1, 0, 1, 1, 1},\nonumber\\
f^{(2)}_{14}&=&\epsilon^2 s_{jk} I^{(2)}_{0, 2, 0, 0, 0, 0, 0, 2, 1},\quad
f^{(2)}_{15}= \epsilon^2 \sqrt{(s_{jk}-4m_f^2)s_{jk}}(2I^{(2)}_{0,2,0,0,0,0,0,1,2}+ I^{(2)}_{0, 2, 0, 0, 0, 0, 0, 2, 1}),\nonumber\\
f^{(2)}_{16}&=&\epsilon^2 m_f^2 s_{jk} I^{(2)}_{1, 1, 0, 0, 0, 0,0, 3, 1},\quad f^{(2)}_{17}=\epsilon^4 s_{jk} I^{(2)}_{1,1,0,0,1,0,0,1,1},\nonumber\\
f^{(2)}_{18}&=&-\epsilon^3 \sqrt{s_{ij}(s_{ij}(s_{jk}-m_f^2)^2-4m_f^2 s_{jk}^2)} I^{(2)}_{1,1,1,0,0,0,0,2,1},\nonumber\\
f^{(2)}_{19}&=&\epsilon^2 \sqrt{s_{ij}s_{jk}(s_{ij}s_{jk}-4m_f^2(s_{ij}+s_{jk}))}(\epsilon I^{(2)}_{ 1, 1, 1, 0, 0, 0, 0, 2, 1}+m_f^2I^{(2)}_{ 1, 1, 1, 0, 0, 0, 0, 3, 1}),\nonumber\\
f^{(2)}_{20}&=&-\epsilon^3 s_{ij}(I^{(2)}_{ 1, 1, 1, -1, 0,0, 0, 2, 1}+m_f^2I^{(2)}_{1,1,1,0,0,0,0,2,1}),\quad
f^{(2)}_{21}=-\epsilon^4 s_{ik} I^{(2)}_{0,1,1,0,1,0,0,1,1},\nonumber\\
f^{(2)}_{22}&=& \epsilon^3 \sqrt{s_{ij}s_{jk}(s_{ij}s_{jk}-4m_f^2(s_{ij}+s_{jk}))}I^{(2)}_{0,1,1,0,1,0,0,1,2},\nonumber\\
f^{(2)}_{23}&=&-\epsilon^3 m_f^2 s_{ik} (I^{(2)}_{0,1,2,0,1,0,0,1,1} + 
  I^{(2)}_{ 0, 2, 1, 0, 1, 0, 0, 1, 1}), \quad
  f^{(2)}_{24}=\epsilon^4 s_{ij}^2 I^{(2)}_{ 1, 1, 1, 0, 1, 
  0, 1, 1, 0},\nonumber\\
f^{(2)}_{25}&=& \epsilon^4 \sqrt{s_{ij}s_{jk}(s_{ij}s_{jk}-4m_f^2(s_{ij}+s_{jk}))}I^{(2)}_{ 1, 1, 0, 0, 1, 0, 1, 1, 1},\nonumber\\
f^{(2)}_{26}&=& \epsilon^4 s_{jk} I^{(2)}_{ 0, 1, 1, 0, 1, 0, 0, 1, 1}+\epsilon^4 s_{ij} I^{(2)}_{ 1, 1, 0, 0, 1, -1, 1, 1,1},\nonumber\\
f^{(2)}_{27}&=& \epsilon^4 \sqrt{(s_{ij}-4m_f^2)s_{ij}}\sqrt{s_{ij}s_{jk}(s_{ij}s_{jk}-4m_f^2(s_{ij}+s_{jk}))}
I^{(2)}_{ 1, 1, 1, 0, 1, 0, 1, 1, 1},\nonumber\\
f^{(2)}_{28}&=& \epsilon^4 \sqrt{(s_{ij}-4m_f^2) s_{ij}}(2 s_{jk}I^{(2)}_{ 1, 1, 0, 0, 1, 0, 1, 1, 1}+s_{ij}I^{(2)}_{ 1, 1, 1, 0, 1, -1, 1, 1, 1}),\nonumber\\
f^{(2)}_{29}&=&-\epsilon^4 s_{ij}\left[I^{(2)}_{ 1, 0, 1, 0, 1, 0, 1, 0, 1}+2I^{(2)}_{1, 1, 0, 0, 1, -1, 1, 1, 1}-I^{(2)}_{1, 1, 1, -1, 1, -1, 1, 1, 1}\right.\nonumber\\
&&\left.+ s_{ij}\left(I^{(2)}_{ 1, 0, 1, 0, 1, 0, 1, 1, 1}-I^{(2)}_{1, 1, 1, 0, 1, -1, 1, 1, 1}\right)+ s_{jk} \left(I^{(2)}_{ 1, 1, 1, 0, 1, 0, 1, 1, 0}-2 I^{(2)}_{1, 1, 0, 0, 1, 0, 1, 1, 1}\right)\right],\label{eq:UTbasis4mf}
\end{eqnarray}
where we have suppressed the arguments $s_{ij},s_{jk},s_{ik}$ in $I^{(2)}_{a_1,\cdots,a_9}$ (defined in eq.\eqref{eq:define-int-2L}) and the arguments of $f_n^{(2)}$ that are $x_{ij}=s_{ij}/m_f^2,x_{jk}=s_{jk}/m_f^2,x_{ik}=s_{ik}/m_f^2$.
It is to note that there are relative sign differences for choices of some of the master integrals as compared to ref.~\cite{Caron-Huot:2014lda}. This is simply due to a different propagator convention $l_i^2 -m_f^2$ instead of $-l_i^2 +m_f^2$. Additionally, we have chosen a different overall normalisation for the integrals as compared to ref.~\cite{Caron-Huot:2014lda}. 
These master integrals satisfy the $\epsilon$-form of differential equation \eqref{eq:DE1}. The non-zero matrix elements of $A^{(2)}$ on the right-hand side of eq.\eqref{eq:DE1} are compactly expressed as the following:
{\small
\begin{eqnarray}
\label{eq:A2ME}
\derive A^{(2)}_{2,1}&=&\derive A^{(2)}_{9,11}=\derive A^{(2)}_{20,11}=\derive A^{(2)}_{28,3}=\derive A^{(2)}_{28,4}=\derive A^{(2)}_{29,5}=\derive A^{(2)}_{29,28}=\derive g_1,\nonumber\\ 
\derive A^{(2)}_{2,2}&=&\derive A^{(2)}_{5,5}=\derive A^{(2)}_{13,13}=\derive g_2,\quad
\derive A^{(2)}_{3,2}=\derive A^{(2)}_{7,1}=\derive A^{(2)}_{13,9}=\derive A^{(2)}_{27,22}=\derive A^{(2)}_{28,10}=-2\derive g_1,\nonumber\\ 
\derive A^{(2)}_{3,3}&=&2\derive g_2,\quad
\derive A^{(2)}_{4,2}=\derive A^{(2)}_{5,3}=\derive A^{(2)}_{5,4}=\derive A^{(2)}_{6,7}=\derive A^{(2)}_{11,9}=\derive A^{(2)}_{13,3}=\derive A^{(2)}_{13,4}=\derive A^{(2)}_{28,20}=-\derive g_1,\nonumber\\ 
\derive A^{(2)}_{6,6}&=&\derive A^{(2)}_{10,10}=\derive g_3,\quad \derive A^{(2)}_{7,6}=6\derive g_1,\quad
\derive A^{(2)}_{7,7}=3\derive g_2-\derive g_3,\quad \derive A^{(2)}_{8,6}=-\frac{3}{4}\derive g_3,\nonumber\\
\derive A^{(2)}_{8,7}&=&\frac{1}{4}\derive g_1,\quad \derive A^{(2)}_{9,2}=\derive A^{(2)}_{12,5}=\derive A^{(2)}_{26,5}=-\frac{1}{2}\derive g_1,\quad
\derive A^{(2)}_{9,9}=\derive A^{(2)}_{12,8}=-\derive g_3,\quad \derive A^{(2)}_{9,10}=\frac{3}{2}\derive g_3,\nonumber\\
\derive A^{(2)}_{10,11}&=&\derive A^{(2)}_{13,8}=\derive A^{(2)}_{24,5}=\derive A^{(2)}_{28,8}=\derive A^{(2)}_{28,9}=\derive A^{(2)}_{28,29}=2\derive g_1,\quad \derive A^{(2)}_{11,2}=\derive A^{(2)}_{12,10}=-\frac{1}{2}\derive g_3,\nonumber\\
\derive A^{(2)}_{11,11}&=&\derive g_2-\derive g_3,\quad
\derive A^{(2)}_{12,6}=-\frac{1}{4}\derive g_3,\quad \derive A^{(2)}_{12,13}=\derive A^{(2)}_{26,13}=\frac{1}{2}\derive g_1,\quad \derive A^{(2)}_{13,6}=-\frac{3}{2}\derive g_1,\nonumber\\ 
\derive A^{(2)}_{13,10}&=&3\derive g_1,\quad
\derive A^{(2)}_{14,14}=\derive g_4,\quad \derive A^{(2)}_{14,15}=\derive g_5,\quad \derive A^{(2)}_{15,1}=2\derive g_5,\quad \derive A^{(2)}_{15,14}=-6\derive g_5,\nonumber\\ 
\derive A^{(2)}_{15,15}&=&\derive g_6,\quad
\derive A^{(2)}_{16,14}=-\frac{3}{4}\derive g_4,\quad \derive A^{(2)}_{16,15}=-\frac{1}{4}\derive g_5,\quad \derive A^{(2)}_{17,14}=-\frac{1}{2}\derive g_4,\quad \derive A^{(2)}_{17,16}=-2\derive g_4,\nonumber\\
\derive A^{(2)}_{18,4}&=&\derive A^{(2)}_{26,18}=\derive g_7,\quad \derive A^{(2)}_{18,9}=\derive A^{(2)}_{29,18}=\derive g_8,\quad \derive A^{(2)}_{18,10}=-\frac{3}{2}\derive g_8,\quad \derive A^{(2)}_{18,11}=\derive g_9,\nonumber\\
\derive A^{(2)}_{18,14}&=&-6\derive g_7+\frac{3}{2}\derive g_8,\quad
\derive A^{(2)}_{18,16}=-8\derive g_7,\quad \derive A^{(2)}_{18,18}=\derive A^{(2)}_{20,4}=\derive g_{10},\quad \derive A^{(2)}_{18,19}=\derive g_{11},\nonumber\\ 
\derive A^{(2)}_{18,20}&=&-3\derive g_{7},\quad
\derive A^{(2)}_{19,2}=\derive A^{(2)}_{25,5}=\derive A^{(2)}_{29,27}=\derive g_{12},\nonumber\\ 
\derive A^{(2)}_{19,4}&=&\derive A^{(2)}_{20,19}=\derive A^{(2)}_{25,12}=\derive A^{(2)}_{25,17}=\derive A^{(2)}_{25,26}=\derive g_{13},\nonumber\\
\derive A^{(2)}_{19,9}&=&\derive A^{(2)}_{25,16}=\derive A^{(2)}_{27,28}=2\derive g_{13},\quad
\derive A^{(2)}_{19,10}=\derive A^{(2)}_{19,14}=\derive A^{(2)}_{19,20}=-\frac{3}{2}\derive g_{13},\nonumber\\
\derive A^{(2)}_{19,11}&=&\derive A^{(2)}_{27,24}=\derive A^{(2)}_{27,14}=2\derive g_{12},\quad \derive A^{(2)}_{19,15}=\derive g_{14},\quad \derive A^{(2)}_{19,16}=\derive A^{(2)}_{27,13}=-2\derive g_{13},\nonumber\\ \derive A^{(2)}_{19,18}&=&\derive A^{(2)}_{25,18}=-\frac{1}{2}\derive g_{11},\quad
\derive A^{(2)}_{19,19}=\derive g_{15},\quad 
\derive A^{(2)}_{20,9}=\derive g_{3}-2\derive g_{4}+2\derive g_{10},\nonumber\\ 
\derive A^{(2)}_{20,10}&=&\derive A^{(2)}_{20,14}=\derive A^{(2)}_{26,10}=-\frac{3}{2}\derive g_{3}+\derive g_{4}-\derive g_{10},\quad \derive A^{(2)}_{20,18}=-\derive g_{7},\nonumber\\
\derive A^{(2)}_{20,20}&=&\derive A^{(2)}_{26,20}=-\derive g_{10},\quad \derive A^{(2)}_{21,6}=-\frac{1}{2}\derive g_{3}+\frac{1}{2}\derive g_{4},\quad \derive A^{(2)}_{21,8}=-2\derive g_{3}+2\derive g_{4},\nonumber\\
\derive A^{(2)}_{21,14}&=&\frac{1}{2}\derive g_{3}-\frac{1}{2}\derive g_{4},\quad
\derive A^{(2)}_{21,16}=2\derive g_{3}-2\derive g_{4},\nonumber\\ \derive A^{(2)}_{21,21}&=&\derive A^{(2)}_{21,23}=\derive A^{(2)}_{26,21}=\derive A^{(2)}_{26,23}=-4\derive g_{4}+4\derive g_{10},\quad \derive A^{(2)}_{22,6}=\derive A^{(2)}_{22,14}=-6\derive g_{13},\nonumber\\ 
\derive A^{(2)}_{22,7}&=&\derive A^{(2)}_{25,13}=\derive A^{(2)}_{27,6}=-\derive g_{12},\quad \derive A^{(2)}_{22,8}=\derive A^{(2)}_{22,16}=-8\derive g_{13},\quad \derive A^{(2)}_{22,15}=2\derive g_{14},\nonumber\\
\derive A^{(2)}_{22,21}&=&\derive A^{(2)}_{22,23}=12\derive g_{13},\quad
\derive A^{(2)}_{22,22}=-2\derive g_{4}+4\derive g_{10}+\derive g_{15},\quad \derive A^{(2)}_{23,6}=\derive A^{(2)}_{23,14}=-\frac{3}{2}\derive g_{4}+\frac{3}{2}\derive g_{10},\nonumber\\
\derive A^{(2)}_{23,8}&=&2\derive g_{3}-4\derive g_{4}+2\derive g_{10},\quad 
\derive A^{(2)}_{23,16}=-2\derive g_{3}+\derive g_{10},\quad 
\derive A^{(2)}_{23,21}=\derive A^{(2)}_{23,23}=6\derive g_{4}-6\derive g_{10},\nonumber\\
\derive A^{(2)}_{23,22}&=&-\frac{1}{2}\derive g_{13},\quad \derive A^{(2)}_{25,10}=\derive A^{(2)}_{25,14}=\derive A^{(2)}_{25,20}=\frac{1}{2}\derive g_{13},\quad \derive A^{(2)}_{25,21}=\derive A^{(2)}_{25,23}=-4\derive g_{13},\nonumber\\
\derive A^{(2)}_{25,25}&=&\frac{2}{3}\derive g_{10}+\frac{1}{3}\derive g_{15},\quad \derive A^{(2)}_{26,6}=-\frac{1}{4}\derive g_{3}+\frac{1}{2}\derive g_{4},\quad \derive A^{(2)}_{26,8}=-\derive g_{3}+2\derive g_{4},\nonumber\\
\derive A^{(2)}_{26,14}&=&\frac{1}{2}\derive g_{4}-\derive g_{10},\quad \derive A^{(2)}_{26,16}=\derive A^{(2)}_{29,17}=2\derive g_{4}-4\derive g_{10},\quad \derive A^{(2)}_{27,3}=-2\derive g_{12},\nonumber\\
\derive A^{(2)}_{27,8}&=&\derive A^{(2)}_{27,10}=\derive A^{(2)}_{27,17}=4\derive g_{12},\quad \derive A^{(2)}_{27,16}=8\derive g_{12},\quad \derive A^{(2)}_{27,19}=\derive A^{(2)}_{28,26}=4\derive g_{1},\nonumber\\
\derive A^{(2)}_{27,21}&=&-12\derive g_{12},\quad \derive A^{(2)}_{27,23}=-8\derive g_{12},\quad \derive A^{(2)}_{27,27}=\derive g_{2}+\derive g_{3}+\derive g_{15},\quad \derive A^{(2)}_{27,29}=-4\derive g_{12},\nonumber\\
\derive A^{(2)}_{28,6}&=&\frac{5}{2}\derive g_{1},\quad \derive A^{(2)}_{28,13}=\derive A^{(2)}_{29,29}=-\derive g_{3}+2\derive g_{10},\quad \derive A^{(2)}_{28,18}=-\derive g_{9},\quad \derive A^{(2)}_{28,21}=\derive A^{(2)}_{28,23}=-4\derive g_{1},\nonumber\\
\derive A^{(2)}_{28,27}&=&-\derive g_{13},\quad \derive A^{(2)}_{28,28}=\derive g_{2}-\derive g_{3}+2\derive g_{4}-2\derive g_{10},\quad \derive A^{(2)}_{29,6}=\frac{3}{2}\derive g_{3}-\derive g_{4},\nonumber\\ 
\derive A^{(2)}_{29,8}&=&6\derive g_{3}-4\derive g_{4},\quad \derive A^{(2)}_{29,10}=2\derive g_{3}-2\derive g_{10},\quad \derive A^{(2)}_{29,14}=\derive A^{(2)}_{29,24}=\derive g_{4}-2\derive g_{10},\nonumber\\
\derive A^{(2)}_{29,16}&=&4\derive g_{4}-8\derive g_{10},\quad \derive A^{(2)}_{29,20}=-\derive g_{3}+2\derive g_{4}-2\derive g_{10},\quad \derive A^{(2)}_{29,21}=-4\derive g_{3}+2\derive g_{4}+4\derive g_{10},\nonumber\\
\derive A^{(2)}_{29,23}&=&-4\derive g_{3}+4\derive g_{4},\quad \derive A^{(2)}_{29,26}=-2\derive g_{3}+4\derive g_{4}-4\derive g_{10}.
\end{eqnarray}
}

\section{Helicity amplitudes of the one-loop $W^\pm$ contribution}
\label{sec:oneloopamp4W}

For completeness, we also present the one-loop helicity amplitudes with $W^\pm$ boson. The one-loop helicity amplitudes with the $W^\pm$ boson and its Goldstone boson from the sum of the $162$ Feynman graphs are:
\begin{eqnarray}
-i\mathcal{M}_{++++}^{(0,0,W)}&=&-12\alpha^2\left[1-4\mathop{\sum_{(i,j,k)=(s,t,u),}}_{(t,u,s),(u,s,t)}{\frac{f_6^{(1,2)}(x_i,x_j,x_k)}{\sqrt{x_ix_j(x_ix_j+4x_k)}}}\right],\nonumber\\
-i\mathcal{M}^{(0,0,W)}_{-+++}&=&-12\alpha^2\left[1+2\left(x_s^{-1}+x_t^{-1}+x_u^{-1}\right)\mathop{\sum_{(i,j,k)=(s,t,u),}}_{(t,u,s),(u,s,t)}{f_4^{(1,2)}(x_i,x_j,x_k)}\right.\nonumber\\
&&\qquad \qquad \qquad \left.-\mathop{\sum_{(i,j,k)=(s,t,u),}}_{(t,u,s),(u,s,t)}{r^{(1)}_8(x_i,x_j,x_k)\frac{f_6^{(1,2)}(x_i,x_j,x_k)}{\sqrt{x_ix_j(x_ix_j+4x_k)}}}\right],\nonumber\\
-i\mathcal{M}^{(0,0,W)}_{--++}&=&-12\alpha^2\left[-1+r_9^{(1)}(x_t,x_u,x_s)\frac{f_2^{(1,1)}(x_t,x_u,x_s)}{\sqrt{x_t(x_t-4)}}+r_9^{(1)}(x_u,x_t,x_s)\frac{f_2^{(1,1)}(x_u,x_t,x_s)}{\sqrt{x_u(x_u-4)}}\right.\nonumber\\
&&-\left(r_{10}^{(1)}(x_s,x_t,x_u)+\frac{5}{3}\right)\left(f_4^{(1,2)}(x_t,x_u,x_s)+f_4^{(1,2)}(x_u,x_s,x_t)\right)\nonumber\\
&&-\frac{4}{3}(x_s-1)(x_s-3)\left(\frac{f_6^{(1,2)}(x_s,x_t,x_u)}{\sqrt{x_sx_t(x_sx_t+4x_u)}}+\frac{f_6^{(1,2)}(x_u,x_s,x_t)}{\sqrt{x_sx_u(x_sx_u+4x_t)}}\right)\nonumber\\
&&\left.-r_{12}^{(1)}(x_s,x_t,x_u)\frac{f_6^{(1,2)}(x_t,x_u,x_s)}{\sqrt{x_tx_u(x_tx_u+4x_s)}}\right],\label{eq:W1Lamp}
\end{eqnarray}
where
\begin{eqnarray}
r_{12}^{(1)}(x_i,x_j,x_k)&=&r_{11}^{(1)}(x_i,x_j,x_k)+\frac{4x_i^2-10x_i-5x_jx_k}{3}.
\end{eqnarray}
Here, we have defined $x_s=\frac{s}{m_W^2},x_t=\frac{t}{m_W^2},$ and $x_u=\frac{u}{m_W^2}$ with $m_W$ being the mass of the $W^\pm$ boson. It is interesting to note that, according to eqs.(\ref{eq:f1Lamp}) and (\ref{eq:W1Lamp}), we have the simple relation $\mathcal{M}_{++++}^{(0,0,W)}/\mathcal{M}_{++++}^{(0,0,f)}=\mathcal{M}_{-+++}^{(0,0,W)}/\mathcal{M}_{-+++}^{(0,0,f)}=-3/(2N_{c,f}Q_f^4)$. However, such a simple relation does not hold for $\mathcal{M}_{--++}$, though the form of $\mathcal{M}^{(0,0,W)}_{--++}$ is very close to $\mathcal{M}^{(0,0,f)}_{--++}$. The one-loop amplitudes provided in ref.~\cite{Bardin:2009gq} are in terms of one-loop scalar integrals rather than a UT basis. Our expressions expressed in terms of a UT basis is however shorter than ref.~\cite{Bardin:2009gq}. The squared amplitude has also been numerically checked with \madloop~\cite{Hirschi:2011pa,Alwall:2014hca} for a few random phase space points.

\section{One- and two-loop helicity amplitudes in the low-energy limit}
\label{sec:ampLElimit}

In this appendix, we collect the one- and two-loop helicity amplitudes in the low energy (LE) limit $m_f^2\gg s,-t,-u$ for a given fermion $f$, which has been derived in ref.~\cite{Martin:2003gb} from the Euler-Heisenberg Lagrangian. We express them in terms of the Mandelstam variables instead of the spinor-helicity formalism presented in ref.~\cite{Martin:2003gb}, with $x_s=s/m_f^2,x_t=t/m_f^2$ and $x_u=u/m_f^2$. Up to a global phase, we have the one-loop helicity amplitudes in the LE limit as:
\begin{eqnarray}
-i\mathcal{M}^{(0,0,f)}_{++++,{\rm LE}}&=&-\frac{1}{15}N_{c,f}Q_f^4\alpha^2\left(x_s^2+x_t^2+x_u^2\right),\nonumber\\
-i\mathcal{M}^{(0,0,f)}_{-+++,{\rm LE}}&=&0,\nonumber\\
-i\mathcal{M}^{(0,0,f)}_{--++,{\rm LE}}&=&\frac{11}{45}N_{c,f}Q_f^4\alpha^2x_s^2.
\end{eqnarray}
The two-loop QCD amplitudes in the same limit are
\begin{eqnarray}
\mathcal{M}^{(1,0,f)}_{++++,{\rm LE}}&=&\frac{25}{4}\frac{\alpha_s}{\pi}C_{F,f}\mathcal{M}^{(0,0,f)}_{++++,{\rm LE}},\nonumber\\
\mathcal{M}^{(1,0,f)}_{-+++,{\rm LE}}&=&0,\nonumber\\
\mathcal{M}^{(1,0,f)}_{--++,{\rm LE}}&=&\frac{1955}{396}\frac{\alpha_s}{\pi}C_{F,f}\mathcal{M}^{(0,0,f)}_{--++,{\rm LE}}.
\end{eqnarray}
The two-loop QED amplitudes can be obtained from the relation
\begin{eqnarray}
\mathcal{M}^{(0,1,f)}_{\lambda_1\lambda_2\lambda_3\lambda_4,{\rm LE}}=(\alpha Q_f^2)/(\alpha_s C_{F,f})\mathcal{M}^{(1,0,f)}_{\lambda_1\lambda_2\lambda_3\lambda_4,{\rm LE}}.
\end{eqnarray}

Similarly, the one-loop helicity amplitudes for the $W^\pm$ boson loop in the LE limit $m_W^2\gg s,-t,-u$ with $x_s=s/m_W^2,x_t=t/m_W^2$ and $x_u=u/m_W^2$ are:
\begin{eqnarray}
-i\mathcal{M}_{++++,{\rm LE}}^{(0,0,W)}&=&\frac{1}{10}\alpha^2\left(x_s^2+x_t^2+x_u^2\right),\nonumber\\
-i\mathcal{M}_{-+++,{\rm LE}}^{(0,0,W)}&=&0,\nonumber\\
-i\mathcal{M}_{--++,{\rm LE}}^{(0,0,W)}&=&\frac{14}{5}\alpha^2x_s^2.\label{eq:LEWamp}
\end{eqnarray}
We have derived eq.\eqref{eq:LEWamp} from the exact-$m_W$-dependent result eq.\eqref{eq:W1Lamp}.

\bibliography{literature} 
\bibliographystyle{utphysM}
\end{document}